\documentclass[aps,reprint,showpacs,superscriptaddress]{revtex4-1}  %

\usepackage{graphicx}
\usepackage{dcolumn}
\usepackage{bm}
\usepackage{hyperref}
\usepackage{definitions}
\usepackage{braket}
\usepackage[figuresleft]{rotating}
\usepackage{booktabs}
\usepackage{adjustbox}
\usepackage[table]{xcolor}
\usepackage{verbatim}

\usepackage[utf8]{inputenc}
\usepackage{float}
\usepackage{amsmath,amssymb,mathtools,bm}
\usepackage{dsfont} 
\usepackage{framed}
\usepackage[bbgreekl]{mathbbol}
\usepackage{slashed}
\usepackage{tikz,tikz-feynman,contour}
\usepackage{multirow}
\usepackage[autostyle]{csquotes}

\newcommand{\dd}{\mathrm{d}}

\begin{document}

\preprint{arXiv:2604.16268}

\title{Experimental prospects for quantum decoherence measurements at  colliders} 

\author{Rafael Aoude}%
\affiliation{Higgs Centre for Theoretical Physics,
School of Physics and Astronomy,\\
The University of Edinburgh, Edinburgh EH9 3JZ, Scotland, UK}%

\author{José Manuel Camacho}%
\affiliation{%
 Instituto de Física Corpuscular (IFIC), CSIC‐Universitat de València, C/ Catedrático José Beltrán 2, 46980 Paterna, Spain
}%

\author{Valentin Durupt}%
\affiliation{Centre for Cosmology, Particle Physics and Phenomenology (CP3), Universit\'e Catholique de Louvain, B-1348 Louvain-la-Neuve, Belgium}%

\author{Guillermo García Mir}
\affiliation{%
 Instituto de Física Corpuscular (IFIC), CSIC‐Universitat de València, C/ Catedrático José Beltrán 2, 46980 Paterna, Spain
}

\author{Fabio Maltoni}
\affiliation{Dipartimento di Fisica e Astronomia, Università di Bologna and INFN, Sezione di Bologna, via Irnerio 46, 40126 Bologna, Italy}%
\affiliation{Centre for Cosmology, Particle Physics and Phenomenology (CP3), Universit\'e Catholique de
Louvain, B-1348 Louvain-la-Neuve, Belgium}%
\affiliation{CERN, Theoretical Physics Department, CH-1211 Geneva 23, Switzerland}

\author{Maria Moreno Llácer}
\affiliation{%
 Instituto de Física Corpuscular (IFIC), CSIC‐Universitat de València, C/ Catedrático José Beltrán 2, 46980 Paterna, Spain
}

\author{Kazuki Sakurai}%
\affiliation{%
Institute of Theoretical Physics, Faculty of Physics, University of Warsaw, Pasteura 5, 02-093, Warsaw, Poland
}

\author{Leonardo Satrioni}%
\affiliation{Centre for Cosmology, Particle Physics and Phenomenology (CP3), Universit\'e Catholique de
Louvain, B-1348 Louvain-la-Neuve, Belgium}%

\author{Marcel Vos}
\affiliation{%
 Instituto de Física Corpuscular (IFIC), CSIC‐Universitat de València, C/ Catedrático José Beltrán 2, 46980 Paterna, Spain
}

\date{\today}

\begin{abstract}
We study the impact of radiation on quantum systems defined by the spins of elementary fermion-antifermion pairs produced at colliders. We present predictions for several processes,  showing that energetic final-state radiation can induce decoherence and significantly reduce the entanglement of quantum systems formed by elementary fermion pairs. We investigate the feasibility of observing this effect experimentally in exclusive samples with energetic radiation. A statistically significant signal can be obtained with current data in associated $pp \rightarrow t\bar{t}(g)$ production at the LHC and in $e^+e^- \rightarrow \tau^{+}\tau^{-}(\gamma)$ production at Belle II. Future electron-positron colliders operated at the $Z$ pole or well above the $t\bar{t}$ production threshold will extend these prospects further.
\end{abstract}

\maketitle

\section{Introduction}

Entanglement is among the most distinctive manifestations of quantum mechanics. When the quantum state of the full system cannot be written as a convex sum of products of states, the system is entangled. Quantum entanglement has been studied extensively in quantum optics, and has been demonstrated in a variety of other physical systems in low energy experiments, see Ref.~\cite{Horodecki:2009zz}.

The time evolution of an isolated quantum system is unitary and deterministic, as described by the Schr\"odinger equation. In realistic situations, however, quantum systems interact with additional degrees of freedom in their environment. The evolution of the subsystem of interest is then no longer unitary and must be described within the framework of open quantum systems~\cite{Davidovich:2014lwo}.
In this framework, interactions with unobserved environmental degrees of freedom lead to decoherence of the quantum system~\cite{Zeh:1970zz,Zurek:1991vd,Schlosshauer:2003zy,Schlosshauer:2019ewh}. Information initially encoded in the subsystem is transferred to the environment, and the reduced density matrix of the subsystem becomes mixed.

Recent collider measurements have established that top-quark pairs produced at the LHC can be entangled~\cite{Afik:2020onf,ATLAS:2023fsd,CMS:2024pts}. Recently, evidence for entanglement in Higgs boson decays was reported by the LHC experiments~\cite{ATLAS:2026hye}. Many other reactions at high-energy colliders are being explored from this perspective~\footnote{In this paper, we work within quantum mechanics, hence the arguments of Refs.~\cite{Abel:1992kz,Bechtle:2025ugc} do not apply to our interpretation.},\cite{Severi:2021cnj,Afik:2022kwm,Aoude:2022imd,Aguilar-Saavedra:2022uye,Fabbrichesi:2022ovb,Afik:2022dgh,Severi:2022qjy,Dong:2023xiw,Aguilar-Saavedra:2023hss,Cheng:2023qmz,Maltoni:2024tul,Aguilar-Saavedra:2024hwd,Maltoni:2024csn,White:2024nuc,Cheng:2024btk,Altomonte:2024upf,Han:2024ugl,Aoude:2025jzc,Barr:2021zcp,Ashby-Pickering:2022umy,Aguilar-Saavedra:2022mpg,Aoude:2023hxv,Barr:2024djo,Bernal:2023ruk,DelGratta:2025qyp,Goncalves:2025mvl,Goncalves:2025xer,Aguilar-Saavedra:2026wuq,Fang:2026ddi}.
The wide range of accessible final states and event topologies opens new opportunities to study quantum correlations in high-energy processes. In collider physics, final-state radiation (FSR) provides a natural place to study the interaction with additional degrees of freedom and decoherence, since the emitted quanta can carry away coherence from the spin system under study. So far, however, most collider analyses have neglected the impact of radiation on these correlations. In Ref.~\cite{Aoude:2025ovu}, the effects of radiation-induced decoherence at colliders were formalized and the associated loss of entanglement was computed. These ideas were subsequently applied  to highly energetic final-state particles and further developed in Ref.~\cite{Gu:2025ijz}.

In this Letter, we present a detailed analytical and numerical study of the impact of energetic final-state radiation on entangled fermion-antifermion pairs. We show that QCD/QED radiation can significantly reduce the entanglement of the corresponding two-spin system. We present standard model predictions for top-quark pair production with energetic final-state QCD radiation and for $e^+e^- \to \tau^+\tau^-(\gamma)$ production, and propose concrete experimental tests at current and future collider facilities to observe this effect. 

\section{Final-state radiation in entangled fermion pairs}
\label{sec:FSRimpact} 

As a first illustrative example, we consider the impact of FSR gluons in top-quark pair production in high-energy electron-positron collisions. The argument applies, with minor modifications, to QED radiation and other fermion-pair-production processes (i.e. $e^+e^- \to \tau^+\tau^- \gamma$), and to $t\bar{t}$ production at the LHC. We quantify entanglement using the concurrence $C[\rho]$: a positive (non-zero) value is a necessary and sufficient condition for entanglement in a two-qubit system~\cite{Wootters:1997id}. At tree level, the unnormalized spin-density matrix ${\bm R}_{\mathrm{LO}}$ is
\begin{align}
    {\bm R}_{\rm LO} = 
    \sum_{\rm in\, dof}|{\cal M}_{0}\rangle  \langle {\cal M}_0| \, ,
\end{align}
where ${\cal M}_{0}$ is the tree-level amplitude, the ket carries the $t\bar t$ spin indices, and we sum over the initial-state degrees of freedom. At a center-of-mass energy of 500~\gev{}, the top quarks are relativistic and the concurrence is about 0.13, with only a mild dependence on the beam polarization~\cite{LinearColliderVision:2025hlt,Altakach:2026fpl,Guo:2026yhz}. The concurrence increases further with increasing center-of-mass energy.

We now consider corrections to this tree-level picture. If one of the top quarks emits a gluon before decaying, as shown in Fig.~\ref{fig:ee2ttbarDiagram}, the entanglement of the $t\bar t$ spin system is modified.

\begin{figure}[htb!]
    \centering
    \tikzfeynmanset{
    every blob={/tikz/fill=blue!30,/tikz/inner sep=2pt,/tikz/minimum size=0.5cm},
    every dot={/tikz/fill=red!30,/tikz/inner sep=2pt}}
    \begin{tikzpicture}
     \begin{feynman}
        \vertex at (-1.0,1.4) (a1) {${e^+}$};
        \vertex at (-1.0,-1.4) (a2) {${e^-}$};
        \vertex at (0,0) (a) ;
        \vertex at (1.75,0) (b);
        \vertex at (3.0,-0.2) (c) {${g(k)}$};
        \vertex at (3.0,1.4) (u) {${t(p_1)}$};
        \vertex at (3.0,-1.4) (d) {${\bar{t}(p_2)}$};
        \vertex[dot,label={above left:\color{red!50}$g_{\rm S}$}] at (2.25,0.58) (u2) {};
        \node (bb) at (0.9,-0.4) {$h,\gamma,Z$};
        \diagram*{
            (a1) -- [anti fermion] (a),
            (a2) -- [fermion] (a),
            (a) -- [photon] (b),
            (b) -- [fermion] (u2) --[fermion] (u),
            (b) -- [anti fermion] (d),
            (u2) -- [photon] (c),
        };
    \end{feynman}
    \end{tikzpicture}
    \caption{Real emission contribution to $e^+e^-\to t{\bar t}g$}
    \label{fig:ee2ttbarDiagram}
\end{figure}%

To describe corrections beyond the Born picture, we organize the spin-density matrix in the form of a next-to-leading-order expansion~\cite{Aoude:2025ovu},
\begin{equation}
{\bm R} = {\bm R}_{\rm LO} + {\bm R}_{\rm NLO}^{\rm virt.} + {\bm R}_{\rm NLO}^{\rm real}\, ,
\label{eq:R}
\end{equation}
where the three terms denote the Born, virtual, and real-emission contributions, respectively. The normalized density matrix is then defined as ${\bm \rho} = {\bm R}/\mathrm{tr}[{\bm R}]$.  While this decomposition provides the natural formal framework, our phenomenological analysis will focus on the real-emission contribution, which captures the effect of real final-state radiation. 


The leading-order result has been discussed extensively in the literature, e.g. in Ref.~\cite{Maltoni:2024csn} and virtual corrections are provided in Ref.~\cite{Aoude:2025ovu}. Here, we focus on the real contribution to the ${\bm R}$-matrix, defined as:
\begin{align}
{\bm R}_{\rm NLO}^{\rm real} =  \sum_{h} |{\cal M}_{0}^{(g),h}\rangle\langle {\cal M}_{0}^{(g),h}| \,,
\end{align}%
where ${\cal M}_{0}^{(g),h}$ is the tree-level amplitude with one gluon emission and the sum is over gluon helicity, reducing it to a $4\times4$ matrix.
The amplitude for the real emission has a $\gamma^\mu$ current, which can be rewritten in terms of a monopole and a dipole contribution, with the latter flipping the spin.
To study how the entanglement depends on the emitted radiation, we introduce the following differential density matrix:
\begin{align}
    \dd \bm\rho = \frac{1}{\mathrm{tr}[{\bm R}]} \left[\left({\bm R}_{\rm LO} + {\bm R}_{\rm NLO}^{\rm virt.}\right)\dd\Phi_2 + {\bm R}_{\rm NLO}^{\rm real}\dd\Phi_3\right] \, .
    \label{eq:NLO}
\end{align}

The quantity defined above is infrared-safe~\cite{Aoude:2025ovu} and remains a density matrix for the $t\bar t$ spin subsystem, differential in the radiation phase space. To probe how the entanglement depends on the energy of the emitted gluon in events with a real emission, we construct an energy-binned density matrix by integrating over the gluon direction at fixed $E_g>0$. The concurrence is then computed from this binned density matrix with the appropriate weights.
This definition is inclusive in the sense that the integration over the unresolved gluon kinematics is performed before evaluating the concurrence.

\begin{figure}[!htb]
    \centerline{
    \includegraphics[width=0.98\linewidth]{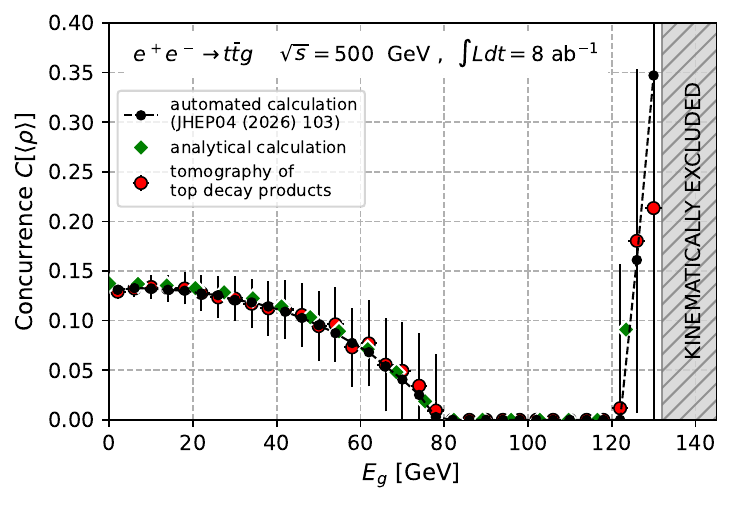}}
    \caption{\leftskip.5em \rightskip.5em The concurrence as a function of the energy of the final-state-radiation gluon in top quark pair production in $e^+e^-$ collisions at $\sqrt{s}=$ 500~\gev. The error bars represent statistical uncertainties, calculated considering all reconstructible final states, with a 60\% acceptance for hadronic modes.
    }
    \label{fig:concurrence_vs_egluon}
\end{figure}%

Figure~\ref{fig:concurrence_vs_egluon} presents the concurrence determined from ${\bm R}_{\rm NLO}^{\rm real}$ as a function of the energy of the FSR gluon. 
Three independent predictions are shown: a semi-analytical calculation~\cite{Aoude:2026} (blue crosses), an automated calculation based on the tools developed in Ref.~\cite{Durupt:2025wuk} (red dashed curve), and a tomographic reconstruction from top-quark decay products in a Monte Carlo sample generated with \MGaMC{}~\cite{Alwall:2014hca} and \MadSpin{}~\cite{Artoisenet:2012st} (red markers with error bars). Excellent agreement is found between the three sets of results.

The concurrence depends strongly on the gluon energy, as shown in Fig.~\ref{fig:concurrence_vs_egluon}. When the gluon is soft its impact on the entanglement of the $t\bar t$ system is small~\cite{Aoude:2025ovu,Gu:2025ijz}, as top-quark spin flips are suppressed. Energetic collinear emissions and higher orders in the soft expansion can flip the top-quark spin. When the gluon carries a sizable fraction of the top-quark momentum the concurrence decreases markedly, reaching zero at around $E_g=80~\gev$. 
Near the kinematic endpoint of the gluon-energy spectrum the concurrence increases again, as a strongly entangled spin-singlet, color-octet state is formed, with the zero-relative-momentum $t\bar t$ system recoiling against the gluon.

The loss of entanglement due to FSR is accompanied by a reduction of the normalized purity, coherence~\cite{Baumgratz:2013ecx} and discord~\cite{Ollivier:2001fdq,Henderson:2001wrr}, as shown in the \enquote{End Matter}, indicating that FSR provokes decoherence of the $t\bar{t}$ system.


Arguably, a complete analysis of the quantum system after FSR emission must consider the spin degrees of freedom of all three particles. For a hypothetical scalar decay $S \rightarrow t\bar{t}g$ with the gluon emitted at a fixed angle (a pure three-qubit state) the tripartite entanglement~\cite{Jin:2022kxb,Sakurai:2023nsc,Horodecki:2025tpn} increases monotonically with $E_g$, as the concurrence of the $t\bar{t}$ system decreases. In the $e^+e^- \rightarrow t\bar{t}g$ process and after integrating over all  angles, the genuine multipartite negativity~\cite{Jungnitsch:2011izf,Hofmann:2014ywl} is zero for all $E_g$, except for a narrow region at the kinematic endpoint.
A complete study of the tri-partite entanglement is found in Ref.~\cite{Goncalves:2026nnx} for the $e^+e^- \rightarrow t \bar{t} Z$ process and will appear in the future~\cite{Sakurai:2026} for the processes considered in this paper.  


\section{Experimental prospects}
\label{sec:Experimental}

In this section, the feasibility of a measurement of quantum decoherence and loss of entanglement is investigated for top-quark pair production in association with a gluon and for $\tau^+\tau^-$ production with an FSR photon, considering present and future collider experiments. The degree of entanglement of the fermion pair is inferred from the tomography of the decay products. The polarization and spin correlations of the entangled fermions are reconstructed from the angular distributions of the decay products, with definitions of the reference frames following Ref.~\cite{Bernreuther:2001rq,Bernreuther:2015yna}. 
Entanglement markers $D$, $D_n$, $D_r$ and $D_k$ are reconstructed from linear combinations of the diagonal elements of the spin correlation matrix: 
\begin{equation}
\begin{split}
  D  & = \frac{1}{3}\left( C_{kk}{+}C_{rr}{+}C_{nn} \right) \,,\\ 
 D_n & = \frac{1}{3}\left( C_{kk}{+}C_{rr} {-} C_{nn} \right) \,,\\ 
 D_r & = \frac{1}{3}\left( C_{kk}{-}C_{rr}{+}C_{nn} \right) \,,\\ 
 D_k & = \frac{1}{3}\left( -C_{kk}{+}C_{rr}{+}C_{nn} \right).
\end{split}
\label{eq:entanglementmarkers}
\end{equation}
If any of these four markers is smaller than -1/3 the two fermions are entangled~\cite{Afik:2020onf}. Also the concurrence can be reconstructed from the tomography. Experimentally, the linear entanglement markers are preferred as they are less prone to bias. We have performed the analysis with the entanglement marker $D_n$ and with the concurrence, and the numerical results for the significance and the conclusions are identical in both cases. 
To simplify comparison with other predictions we report concurrence results in this Letter.

The loss of entanglement caused by high-energy FSR emissions is revealed by a differential measurement of the degree of entanglement versus the energy of the final-state radiation. For each interval of FSR energy a full characterization of the spin correlations of the quantum system can be performed. In the following, we consider a simple analysis, performing two entanglement measurements~\footnote{We take loss of entanglement as a proxy for decoherence; a much more complete diagnosis of the quantum state is possible experimentally, with a full characterization of the spin correlation matrix, computing purity, discord, etc.} The \textit{reference} measurement is performed in a \textit{control sample} with no or soft additional radiation and compared to a \textit{decoherence} measurement in a \textit{signal sample} with an energetic emission. The ratio $R = C[\rho]_{\rm signal}/C[\rho]_{\rm control}$ of the measurements in the two samples signals decoherence whenever $R$ differs significantly from 1. Statistical uncertainties are estimated on generator-level predictions and scaled to realistic operation scenarios of the different colliders. The projections are compared to existing measurements and more detailed projections to assess the practical feasibility in realistic experimental conditions. A complete assessment of systematic uncertainties remains, however, beyond the scope of this Letter.

\begingroup \renewcommand{\arraystretch}{1.5}
\begin{table*}[htb!]
\begin{tabular}{|cccccccccc|}
\hline
\multicolumn{1}{|c|}{\multirow{2}{*}{Machine}} & \multicolumn{1}{c|}{\multirow{2}{*}{\begin{tabular}[c]{@{}c@{}}$\sqrt{s}$\\ $\left[   \text{TeV} \right]$\end{tabular}}} & \multicolumn{1}{c|}{\multirow{2}{*}{\begin{tabular}[c]{@{}c@{}}$\int \mathcal{L}\, d t $\\ $  \left[ \text{fb}^{-1} \right]$\end{tabular}}} & \multicolumn{2}{c|}{$t\bar{t}$} & \multicolumn{3}{c|}{$t\bar{t}+jet$} & \multicolumn{1}{c|}{\multirow{2}{*}{R  = $\frac{C_{t \bar{t}+jet}}{C_{t \bar{t}}}$}} & \multirow{2}{*}{$\mathcal{S} = \frac{R-1}{\sigma_{R}}$} \\ \cline{4-8}
\multicolumn{1}{|c|}{} & \multicolumn{1}{c|}{} & \multicolumn{1}{c|}{} & \multicolumn{1}{c|}{\begin{tabular}[c]{@{}c@{}}$\#$ \small events\end{tabular}} & \multicolumn{1}{c|}{C$\left[ \rho \right]$} & \multicolumn{1}{c|}{Cut} & \multicolumn{1}{c|}{\begin{tabular}[c]{@{}c@{}}$\#$ \small events\end{tabular}} & \multicolumn{1}{c|}{C$\left[ \rho \right]$} & \multicolumn{1}{c|}{} &  \\ \hline
\multicolumn{10}{|c|}{$e^+e^-$ collisions} \\ \hline
\multicolumn{1}{|c|}{\multirow{2}{*}{\begin{tabular}[c]{@{}c@{}}linear   $e^+ e^-$ \\ LCF \cite{LinearColliderVision:2025hlt}\end{tabular}}} & \multicolumn{1}{c|}{0.5} & \multicolumn{1}{c|}{8} & \multicolumn{1}{c|}{4.38$\cdot 10^6$} & \multicolumn{1}{c|}{0.133} & \multicolumn{1}{c|}{$p_T^g > 30$ GeV} & \multicolumn{1}{c|}{0.14$\cdot 10^6$} & \multicolumn{1}{c|}{0.04} & \multicolumn{1}{c|}{$0.31 \pm \stackrel{\text{dilep.}}{0.23} / \stackrel{\text{all ch.}}{0.11}$} & $\stackrel{\text{dilep.}}{3.0 \sigma} / \stackrel{\text{all ch.}}{8.2 \sigma}$ \\ \cline{2-10} 
\multicolumn{1}{|c|}{} & \multicolumn{1}{c|}{1.0} & \multicolumn{1}{c|}{8} & \multicolumn{1}{c|}{1.32$\cdot 10^6$} & \multicolumn{1}{c|}{0.249} & \multicolumn{1}{c|}{$p_T^g > 120$ GeV} & \multicolumn{1}{c|}{0.10$\cdot 10^5$} & \multicolumn{1}{c|}{0.08} & \multicolumn{1}{c|}{$ 0.33 \pm \stackrel{\text{dilep.}}{0.15}  / \stackrel{\text{all ch.}}{0.07}$} & $\stackrel{\text{dilep.}}{4.6 \sigma} / \stackrel{\text{all ch.}}{12 \sigma}$ \\ \hline
\multicolumn{10}{|c|}{$pp$ collisions (Fiducial region:   $|\cos\Theta_{t}^{t\bar{t}RF}|<0.4, m_{t\bar{t}} > 800$ GeV)} \\ \hline
\multicolumn{1}{|c|}{LHC Run   2} & \multicolumn{1}{c|}{13} & \multicolumn{1}{c|}{140} & \multicolumn{1}{c|}{0.57$\cdot 10^6$} & \multicolumn{1}{c|}{0.405} & \multicolumn{1}{c|}{$p_T^j > 250$ GeV} & \multicolumn{1}{c|}{0.06$\cdot 10^6$} & \multicolumn{1}{c|}{0.12} & \multicolumn{1}{c|}{$  0.29 \pm \stackrel{\text{dilep.}}{0.12} / \stackrel{\text{all ch.}}{0.04} $} & $\stackrel{\text{dilep.}}{5.8 \sigma} / \stackrel{\text{all ch.}}{16 \sigma}$ \\ \hline
\multicolumn{1}{|c|}{HL-LHC~\cite{ATLAS:2025eii}} & \multicolumn{1}{c|}{14} & \multicolumn{1}{c|}{3000} & \multicolumn{1}{c|}{14.9$\cdot 10^6$} & \multicolumn{1}{c|}{0.431} & \multicolumn{1}{c|}{$p_T^j > 250$ GeV} & \multicolumn{1}{c|}{1.77$\cdot 10^6$} & \multicolumn{1}{c|}{ 0.10} & \multicolumn{1}{c|}{$  0.22 \pm \stackrel{\text{dilep.}}{0.06} / \stackrel{\text{all ch.}}{0.02}$} & $\stackrel{\text{dilep.}}{12 \sigma} / \stackrel{\text{all ch.}}{34 \sigma}$ \\ \hline
\end{tabular}
    \caption{Prospected statistical significance for the loss of entanglement between $t\bar{t}$ production and $t\bar{t}$ production in association with an energetic jet, from $e^+ e^-$ and pp collisions considering the di-lepton decay mode (dilep.) and a combination of all final states (all ch.), assuming a 60~\% down-type quark acceptance. Concurrence predictions are provided for the inclusive $t \bar{t}$ sample and with an optimized $p_T$ cut for $t \bar{t} + j$.} 
    \label{tab:prospects_ttgluon}
\end{table*}

\noindent\textbf{Top quark  production at high-energy colliders.}
To assess the prospect of an observation of decoherence, we consider operation of a lepton collider at centre-of-mass energies ranging from 500~\gev{} to 3~\tev{}, as foreseen at the Linear Collider Facility at CERN~\cite{LinearColliderVision:2025hlt} or at a muon collider~\cite{InternationalMuonCollider:2025sys}.

In Table~\ref{tab:prospects_ttgluon} results are  presented for inclusive $e^{+}e^{-} \rightarrow t \bar{t}$ production and for an exclusive $e^{+}e^{-} \rightarrow t \bar{t} g$ sample, where the top quark pair is accompanied by an energetic gluon. The inclusive sample is strongly entangled.
The concurrence in the sample with an energetic FSR gluon radiation is significantly reduced. We optimized the \pt{} cut for maximal statistical significance in the di-lepton channel. 
A comparison of the concurrence measurements in the two samples gives access to quantum decoherence. The statistical significance of the ratio $R = C_{t\bar{t}g}/C_{t\bar{t}}$ of the concurrence measurements is presented in the last column of Table~\ref{tab:prospects_ttgluon}. 

Final states with two prompt leptons ($e^+e^-$, $e^\pm \mu^\mp$ and $\mu^+\mu^-$ signatures, but disregarding $\tau$-lepton final states) provide straightforward polarimetry with a spin analyzing power $\sim$ 1. An analysis of these di-lepton events has enough statistical power to claim evidence for quantum decoherence. An analysis of the much more abundant events with one or two hadronic top quark decays requires identification of the jet formed by the down-type quark. We assign an efficiency of 60~\% (per top decay) to tag the down-type quark, based on the work in Refs.~\cite{Dong:2023xiw,Dong:2024xsb,Dong:2024xsg}. Under this conservative assumption, the inclusion of the lepton+jets and fully hadronic channels is expected to boost the statistical significance to well over the conventional 5$\sigma$ bound.

\begin{figure}[h!]
\centerline{
\includegraphics[width=0.98\linewidth]{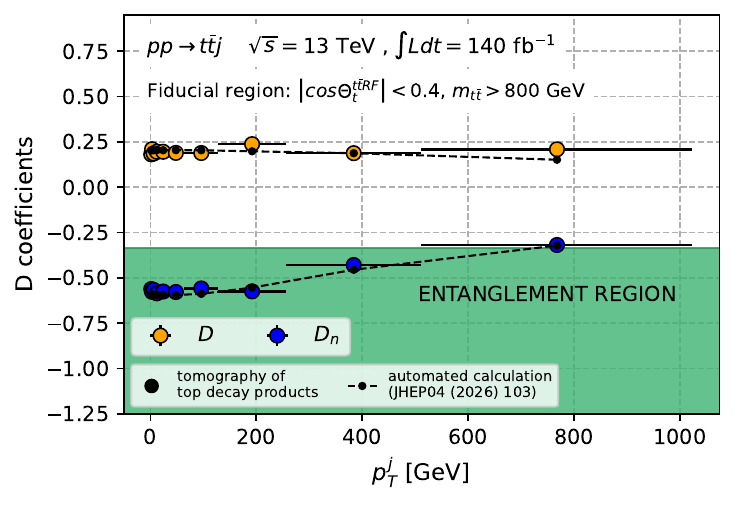}}
\caption{\leftskip.5em \rightskip.5em The entanglement markers $D$ (\textcolor{orange}{orange}) and $D_n$ (\textcolor{blue}{blue}) in $pp \rightarrow t \bar{t} j$ production in a \enquote{boosted top} fiducial region at $\sqrt{s} =$ 13~\tev{}, as a function of the \pt{} of the additional jet. The error bars represent statistical uncertainties, calculated considering all reconstructible final states, with a 60\% acceptance for hadronic modes.}
\label{fig:lhc_markers_vs_jet_pt}
\end{figure}


A large sample of events with high-\pt{} top quarks has been collected by ATLAS and CMS at the LHC~\cite{ATLAS:2022xfj}. The CMS experiment has observed entanglement~\cite{CMS:2024zkc} in events with $m_\ttbar >$ 800~\gev{} and $|\cos \Theta_t^{t\bar{t}RF}| <$ 0.4, where $m_\ttbar$ is the invariant mass of the top quark pair and $\Theta_t^{t\bar{t}RF}$ is defined as the top quark scattering angle in the \ttbar{} center-of-mass
frame. In the boosted regime, as in the $e^+e^- \rightarrow t\bar{t}$ process, the $t\bar{t}$ pair is produced in an entangled state at the LHC and FSR gluons lead to decoherence. Additional effects play a role in $pp$ collisions; there is an important dilution by $gg\rightarrow t\bar{t}$ production and from ISR gluons. The broad $m_{t\bar{t}}$ distribution smears out the effect. To estimate their effects, events are generated for the $pp \rightarrow t \bar{t} j$ process (with $j$ any additional parton) in the fiducial region of Ref.~\cite{CMS:2024zkc}. The pseudo-rapidity of the additional jet is limited to the detector coverage $| \eta_j | < $ 5. 

The expected dependence of the entanglement markers $D$ and $D_n$ 
on the \pt{} of the additional jet is shown in Fig.~\ref{fig:lhc_markers_vs_jet_pt}. While the shape differs in important aspects from the $e^+e^-$ case, a clear dependence of the entanglement markers on jet \pt{} is visible. $D_n$, in particular, increases steadily with jet \pt{}.

The results in the lower half of Table~\ref{tab:prospects_ttgluon} show that the statistical power of the LHC experiments exceeds that of the $e^+e^-$ collider. The entries labelled \enquote{LHC run 2} indicate that observation of a significant loss of entanglement is possible with a single experiment and the existing run 2 data, in the experimentally straightforward di-lepton channel. The lepton+jets and all-hadronic channels enhance the significance to a comfortable 20$\sigma$ per experiment. With run 3 of the LHC, the $pp$ collision data has more than doubled and the ATLAS and CMS experiments will collect several million events each in the high-luminosity program of the LHC, boosting the sensitivity still further. 

\hspace{-3cm} \begin{table*}[htb!] 
\begin{tabular}{|c|c|c|cc|ccc|cc|c|}
\hline
\multirow{2}{*}{Machine} & $\sqrt{s}$ & $\int \mathcal{L}\, d t$ & \multicolumn{2}{c|}{$\tau^+ \tau^-$} & \multicolumn{3}{c|}{$\tau^+ \tau^- \gamma$} & \multicolumn{2}{c|}{\multirow{2}{*}{$R = \frac{C_{\tau^+ \tau^- \gamma}}{C_{\tau^+   \tau^-}}$}} & \multirow{2}{*}{$\mathcal{S} = \frac{R-1}{\sigma_{R}}$} \\ \cline{4-8}
 & $\left[ \text{GeV} \right]$ & $\left[ \text{ab}^{-1} \right]$ & \multicolumn{1}{c|}{$\#$ events} & C$\left[ \rho \right]$ & \multicolumn{1}{c|}{Cut} & \multicolumn{1}{c|}{$\#$ events} & C$\left[ \rho \right]$ & \multicolumn{2}{c|}{} &  \\ \hline
\multicolumn{11}{|c|}{$e^+e^-$ collisions (Restriction to   $|\cos\Theta_{t}^{t\bar{t}RF}|<0.4)$} \\ \hline
\multirow{2}{*}{Belle II} & \multirow{2}{*}{10.58} & $\stackrel{\text{2025 }}{0.5}$ & \multicolumn{1}{c|}{$\stackrel{\text{2025}}{1.48\cdot10^{6}} $} & \multirow{2}{*}{0.727} & \multicolumn{1}{c|}{\multirow{2}{*}{$p_T^\gamma > 2$ GeV}} & \multicolumn{1}{c|}{$\stackrel{\text{2025}}{0.02\cdot10^{6}}$} & \multirow{2}{*}{0.235} & \multicolumn{1}{c|}{\multirow{2}{*}{$0.32$}} & $\pm \stackrel{\text{2025}}{0.01}$ & $\stackrel{\text{2025}}{28 \sigma}$ \\ \cline{3-4} \cline{7-7} \cline{10-11} 
 &  & $\stackrel{\text{Target }}{50}$ & \multicolumn{1}{c|}{$\stackrel{\text{Target}}{140\cdot10^{6}}$} &  & \multicolumn{1}{c|}{} & \multicolumn{1}{c|}{$\stackrel{\text{Target}}{2.14\cdot10^{6}}$} &  & \multicolumn{1}{c|}{} & $ \pm \stackrel{\text{Target}}{0.001}$ & $\stackrel{\text{Target}}{280 \sigma}$ \\ \hline
\multirow{2}{*}{Z-pole} & \multirow{2}{*}{91.20} & $\stackrel{\text{Giga-Z }}{0.1} $ & \multicolumn{1}{c|}{$\stackrel{\text{Giga-Z}}{0.72 \cdot 10^{6}} $} & \multirow{2}{*}{0.858} & \multicolumn{1}{c|}{\multirow{2}{*}{$p_T^\gamma > 30$ GeV}} & \multicolumn{1}{c|}{$\stackrel{\text{Giga-Z }}{< 0.01 \cdot   10^{6}} $} & \multirow{2}{*}{0.102} & \multicolumn{1}{c|}{\multirow{2}{*}{$0.12$}} & $ \pm \stackrel{\text{Giga-Z }}{0.05} $ & $\stackrel{\text{Giga-Z }}{16 \sigma}$ \\ \cline{3-4} \cline{7-7} \cline{10-11} 
 &  & $ \stackrel{\text{Tera-Z }}{300}$ & \multicolumn{1}{c|}{$\stackrel{\text{Tera-Z}}{216   \cdot 10^{6}}$} &  & \multicolumn{1}{c|}{} & \multicolumn{1}{c|}{$\stackrel{\text{Tera-Z }}{0.86 \cdot 10^{6}}$} &  & \multicolumn{1}{c|}{} & $ \pm \stackrel{\text{Tera-Z }}{0.001}$ & $\stackrel{\text{Tera-Z }}{900 \sigma}$ \\ \hline
 \end{tabular}
 \caption{The concurrence in $\tau^+\tau^-$ and $\tau^+ \tau^-  \gamma$ production at the Belle II experiment and a future $e^+e^-$ collider at the $Z$-pole~\cite{LinearColliderVision:2025hlt,FCC:2025lpp,CEPCStudyGroup:2018ghi}.   The statistical significance of the decoherence measurement is indicated for operating scenarios of existing or planned facilities.}   
 \label{tab:prospects_tautauphoton}
\label{tab:prospects_tautauphoton}
\end{table*} 

\noindent\textbf{$\tau$-lepton production at Belle II and the Z-pole.}
High-luminosity electron-positron colliders operated as a $B$-hadron or $Z$-boson factories are expected to produce samples of 10$^8$ to 10$^{11}$ $\tau^+ \tau^-$ pairs in entangled states~\cite{Ehataht:2023zzt,Fabbrichesi:2024wcd}. The prospects of the Belle II experiment and the experiments in Giga-Z and Tera-Z runs of linear and circular electron-positron collider projects ~\cite{LinearColliderVision:2025hlt,CEPCStudyGroup:2018ghi,FCC:2025lpp} are assessed here. Also an upgraded $\tau$/charm factory~\cite{Ai:2025xop} at $\sqrt{s} \sim $ 4~\gev{} presents excellent opportunities~\cite{Yang:2026uwu}.

Following the prospect study for Belle II of Ref.~\cite{Ehataht:2023zzt} the one-prong $\tau^- \rightarrow \pi^- \nu_\tau$ decay, with a branching fraction of about 10\% and spin analyzing power $\sim 1$, is simulated at $\sqrt{s}=$ 10.579~\gev and $\sqrt{s} = m_Z$. 

Table~\ref{tab:prospects_tautauphoton} presents the results for the concurrence of $e^+ e^- \rightarrow \tau^+ \tau^-$ and $e^+ e^- \rightarrow \tau^+ \tau^- \gamma$ samples. The $\pt$ cut on the photon is set to 2~\gev{} at the $B$-factory and raised to 30~\gev{} at the $Z$-pole, to maximize the significance. Only $\tau^+ \tau^-$ pairs in the central region of the detector are considered ($|\cos \Theta_\tau^{RF} | < 0.4$, where $\Theta_\tau^{RF}$ is the $\tau^-$ scattering angle in the center-of-mass rest frame), as in Ref.~\cite{Ehataht:2023zzt}. In both cases, the concurrence is high in the inclusive $\tau^+\tau^-$ sample and drastically reduced in an exclusive sample where an energetic photon is tagged. The statistical significance of the ratio of the two measurements is expected to reach well beyond 5$\sigma$ with the \mbox{Belle II} data collected to date, and even with the limited GigaZ program foreseen at linear colliders. The data samples envisaged in the target integrated luminosity of Belle II and the Tera-Z program of circular colliders yield a statistical  significance $\gg 5\sigma$. The abundant statistics allows for a fine-grained differential measurement, characterizing the quantum state in narrow bins of the FSR energy. These measurements can map out the impact of FSR on the spin correlation matrix, the degree of entanglement, the purity and coherence and discord of the quantum system with excellent precision.

\section{Summary and Discussion}
In this Letter, we have studied the impact of energetic final-state radiation on an entangled two-fermion system with analytic calculations and simulated samples. We find that entangled fermion pairs produced at colliders show significant decoherence in events with energetic final-state radiation.  Several collider experiments can observe this effect. A statistically significant demonstration is possible already today with the data collected in $\tau^+\tau^- (\gamma)$ production at the Belle II experiment and with the samples of boosted $t\bar{t} (g)$ events collected by the ATLAS and CMS experiments at the Large Hadron Collider. 
Experiments at a future electron-positron collider can study this effect in $\tau^+ \tau^- (\gamma)$ production at the $Z$-pole and in $t\bar{t} (g)$ production for $\sqrt{s} > 500~\gev$.

It should be emphasized that these projections consider only statistical uncertainties. Measurements of spin correlations and entanglement in boosted top quark pair production at the LHC and precision measurements of $\tau$-leptons at Belle II have demonstrated excellent control of the systematic uncertainties~\cite{CMS:2024zkc, Belle-II:2023izd}, that should be adequate for observation of decoherence. A follow-up analysis using full detector simulation should be performed to refine these projections.

This proposal establishes a
new avenue for experimental studies of quantum decoherence in
high-energy collisions. The measurements at colliders proposed here complement existing measurements in dedicated low-energy experiments
\footnote{Searches for decoherence in collider experiments such as CPLEAR~\cite{Bertlmann:1999np} and in neutrino experiments~\cite{ICECUBE:2023gdv,KM3NeT:2024jji,deGouvea:2021uvg} have yielded null results so far.},
\cite{Brune:1996zz,Arndt:1999kyb,Schlosshauer:2019ewh}. The decoherence due to final-state radiation studied in this Letter can be traced to interactions described by the standard model, providing a concrete setting in which this effect can be investigated quantitatively at high-energy colliders.

A future extension of this work will further study the perturbative split between the quantum system and the environment. The $2\to3$ process can be interpreted either as the $2\to2$ system coupled to an environment at order $\alpha_s$, and hence as an open quantum system, or as a closed three-particle quantum system that first arises at order $\alpha_s$. Investigating the correlations between the fermion-pair spins and the polarization of the final-state radiation at different perturbative orders may help clarify how unobserved degrees of freedom affect the quantum state of the observed subsystem.

\begin{acknowledgments}
We thank Alan Barr, Federica Fabbri, Juan Ramón Muñoz de Nova, Martin White and Laura Zani for discussions. We thank CERN and the LHC physics center for their hospitality. The IFIC group is supported by various projects of the Spanish Ministry and State Research Agency (PID2024-156321NB-I00, PID2024-158190NBC21 and CNS2024-154901), of the regional government Generalitat Valenciana (CIPROM/2021/073, CIPROM/2022/70 and pre-doctoral grant CIACIF/2024/194), and the Severo Ochoa excellence programme. R.A. is supported by UK Research and
Innovation (UKRI) under the UK government’s Horizon Europe Marie Sklodowska Curie funding guarantee grant [EP/Z000947/1]. This research is partially supported by the IISN convention 4.4517.08, ``Theory of fundamental interactions.''

\end{acknowledgments}

\bibliography{decoherence_prl}

\clearpage
\onecolumngrid

\newpage
\begin{center}
  \textbf{\large End Matter}\\[.2cm]
  \vspace{0.05in}
  {Rafael Aoude, Jos\'e Manuel Camacho,  Valentin Durupt, Fabio Maltoni, Maria Moreno Ll\'acer, Kazuki Sakurai, Leonardo Satrioni, and Marcel Vos}
\end{center}

\setcounter{equation}{0}
\setcounter{figure}{0}
\setcounter{page}{1}
\makeatletter
\renewcommand{\theequation}{S\arabic{equation}}

In this \enquote{End Matter} we present predictions for several quantum observables to characterize the quantum systems discussed in the paper. All results are based on Monte Carlo samples for the $e^+e^- \rightarrow t\bar{t} $ and $e^+e^- \rightarrow t \bar{t} g$ processes generated with MadGraph5\_aMC@NLO. The events in the Les Houches Accord format are further processed with the "reweight" option of Ref.~\cite{Durupt:2025wuk} to generate the spin correlation matrix, and with MadSpin to simulate the $t \rightarrow Wb \rightarrow \mu \nu_\mu b$ decay chains. The elements of the spin correlation matrix are then reconstructed from the angular distributions of the muons in the parent top quark rest frame. We use the helicity basis and matrices and polarization vectors are expressed in the $\{n, r, k\}$ basis, following the convention
of Ref.~\cite{Bernreuther:2001rq,Bernreuther:2015yna}. In predictions for $e^+e^-$ collisions we do not apply the sign flips adopted by Bernreuther for symmetric LHC collisions.  Note that, following the conventions of Ref.\cite{CMS:2024zkc}, the sign of the spin correlation matrix elements from tomography is opposite with respect to those obtained using the automated calculation of Ref. \cite{Durupt:2025wuk}. 

\vspace{1em}
\begin{figure} [!hbt]
\centerline{
    \includegraphics[width=0.6\linewidth]{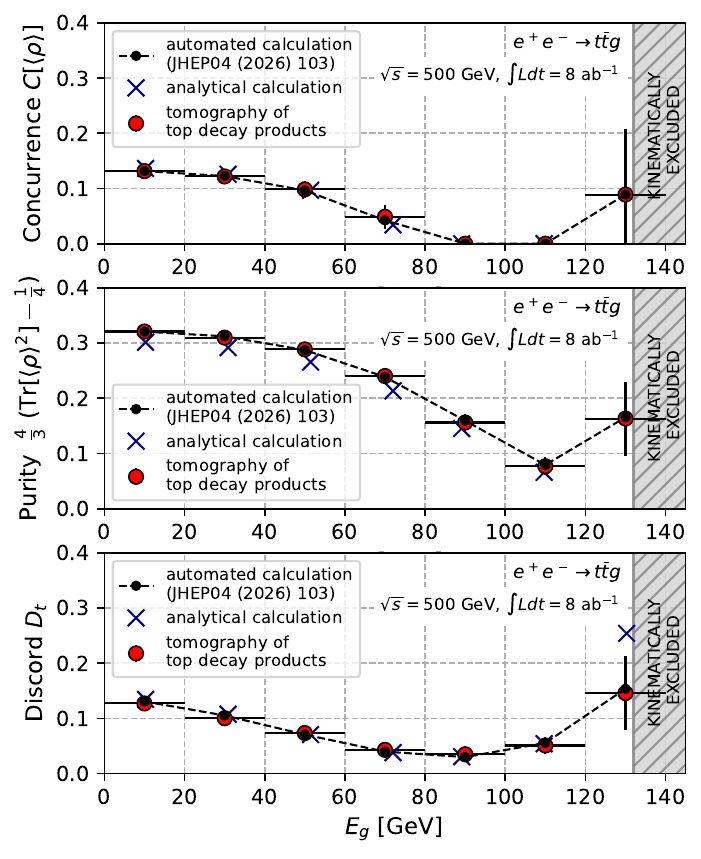}}
    \caption{The concurrence (upper panel), purity (central panel) and discord (lower panel) as a function of gluon energy in the $e^+e^- \rightarrow \gamma^* / Z^* \rightarrow t\bar{t} g$ process at $\sqrt{s}=$ 500~\gev. Automated predictions are shown as dashed curves with black markers, analytical predictions as blue crosses, and tomography results as large, red markers with error bars corresponding to the expected statistical uncertainty with an integrated luminosity of 8~\iab{}. The quantum observables are defined in the text.}
    \label{fig:concurrence_purity_discord}
\end{figure}

In Fig.~\ref{fig:concurrence_purity_discord}, the dependence of the concurrence on the gluon energy is compared to the discord and purity. The purity is calculated as: 
\begin{equation} 
\frac{d}{d-1}({\rm Tr}[\rho^2] - \frac{1}{d}),
\end{equation}
where the dimension $d=4$. For a pure state, this indicator yields 1. 
An intuitive measure of the coherence on the $t\bar{t}$ system is estimated as the sum of off-diagonal elements of the density matrix $\rho$. 
The $l_1$ norm~\cite{Baumgratz:2013ecx} (not shown on Fig.~\ref{fig:concurrence_purity_discord}) is defined as:
\begin{equation}
C_l = \sum_{i \neq j} | \rho_{ij}|
\end{equation}
Instead, we plot quantum discord, a mutual-information-based measure of the quantumness of correlations~\cite{Ollivier:2001fdq,Henderson:2001wrr}. For a two-qubit system, discord is defined as:
\begin{equation}
D_t = S(\rho_{\bar{t}}) - S(\rho) + \min_{\hat{n}}{[p_{\hat{n}} S(\rho_{\hat{n}}) + p_{ - \hat{n}} S(\rho_{- \hat{n}})]},
\end{equation}
where the reduced density matrix $\rho_{\bar{t}} = {\mathrm{Tr}}_t \rho$ is obtained from the density matrix $\rho$ of the the $t\bar{t}$ system by tracing out the top quark. $S(\rho) = - \mathrm{Tr} \rho \log_{2} \rho$ represents the Von Neumann entropy. The Bloch vector $\hat{n}$ is varied to find the minimum discord.   
Discord is especially well-suited to characterize the effect of decoherence~\cite{Zurek_2025} and avoids the basis-dependence of coherence. Both quantities 
show a similar trend to purity and concurrence, with a sharp decrease to reach to a minimum for $E_g $ between 80~\gev{} and 120~\gev{}, and a strong increase at the kinematic endpoint.

\vspace{1em}
\begin{figure} [!hbt]
\centerline{
\includegraphics[width=0.6\linewidth]{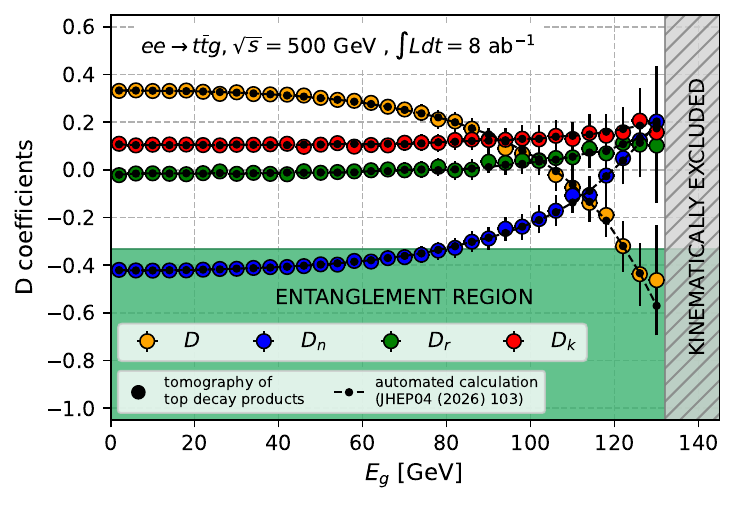}}
    \caption{The value of the entanglement markers $D$, $D_n$, $D_k$ and $D_r$ (see text for a description) of the $t\bar{t}$ system in top quark pair production in association with a gluon ($e^+e^- \rightarrow t\bar{t} g$ production), at $\sqrt{s}=$ 500~\gev{}, as a function of the energy of the FSR gluon. The continuous lines  correspond to a direct determination using the code of Ref~\cite{Durupt:2025wuk}, while the markers have been obtained with a tomography of $t \rightarrow W b \rightarrow l \nu_l b$ decays simulated with MadSpin. }
    \label{fig:D_markers}
\end{figure}

Further insight into the impact of the FSR gluon emission on the degree of entanglement can be gained from Fig.~\ref{fig:D_markers}, that shows the evolution with $E_g$ of four particular linear combinations of the diagonal spin correlation matrix elements of Eq.~\ref{eq:entanglementmarkers}.
In $e^+ e^- \rightarrow t\bar{t}$ production and in events with a low-energy gluon, the top-quark pair is produced primarily in a spin triplet state and only the $D_n$ marker reaches below the entanglement limit of -1/3. For events where the $t\bar{t}$ pair recoils against a high-energy gluon a spin-singlet configuration appears, and the $D$ marker sensitive to singlet states acquires more and more negative values. The region of minimal concurrence corresponds to a "mixed" sample where singlet and triplet states co-exist. This incoherent mixture of states leads to (classical) decoherence of the ensemble of events.



\clearpage
\appendix*


\section{ Reference results for spin polarization, correlations and full density matrix reconstruction}

In this Appendix, we provide, for reference, the numerical results obtained for the polarization vectors $B$ and the spin correlation matrices $C$ determined using tomography, alongside the density matrices $\rho$ reconstructed with these values. These are expressed in the $\{n,r,k\}$ basis, following the conventions of the references~\cite{Bernreuther:2001rq, Bernreuther:2015yna}. Note that the use of the $sign(y_r)$ factor introduced in this reference is reserved to the $pp$ collision samples. In addition, $B^{-}$ refers to the particle polarization vector, while $B^{+}$ refers to the antiparticle polarization vector. The predictions are obtained from Monte Carlo samples, generated with \textsc{MadGraph5\_aMC@NLO}. 

For $t\bar{t}$ production, statistical uncertainties for B and C values are determined considering all decay channels reconstructible at detector level, with a 60\% acceptance in both hadronic channels,  for the emulation of light quarks reconstruction efficiencies. For $\tau^+\tau^-$ production, only the single-prong decay channel is considered, with luminosities of L = 50 ab$^{-1}$ for Belle II and L = 300 ab$^{-1}$ for the FCC-ee Z-pole operation. FSR samples include the identical $p_T$ cut applied for providing with the significance results in Tables \ref{tab:prospects_ttgluon} and \ref{tab:prospects_tautauphoton}. 

Additionally, in this Appendix we present analogous results to Tables \ref{tab:prospects_ttgluon} and \ref{tab:prospects_tautauphoton} alternatively using the $D_n$. Significances obtained using this method are in total agreement with the results discussed in the main part of the study.  

\subsection*{\large Inclusive $\rho_{t\bar{t}}$ reconstruction for $\boldsymbol{e^{+}e^{-} \rightarrow t \bar{t}}$}

\[
\begin{array}{l}
\boldsymbol{e^{+}e^{-} \rightarrow t\bar{t} \quad (\sqrt{s}=500~\textbf{GeV}):}
\end{array} \hfill
\rho_{t\bar{t}} =
\left[\begin{array}{cccc}
\cellcolor[RGB]{208,147,131}\makebox[2.8em][c]{0.323} &
\cellcolor[RGB]{254,254,255}\makebox[2.8em][c]{-0.001} &
\cellcolor[RGB]{255,255,255}\makebox[2.8em][c]{-0.001} &
\cellcolor[RGB]{147,153,229}\makebox[2.8em][c]{-0.153} \\[2pt]
\cellcolor[RGB]{254,254,255}\makebox[2.8em][c]{-0.001} &
\cellcolor[RGB]{232,194,178}\makebox[2.8em][c]{0.087} &
\cellcolor[RGB]{232,195,179}\makebox[2.8em][c]{0.086} &
\cellcolor[RGB]{255,255,255}\makebox[2.8em][c]{0.000} \\[2pt]
\cellcolor[RGB]{255,255,255}\makebox[2.8em][c]{-0.001} &
\cellcolor[RGB]{232,195,179}\makebox[2.8em][c]{0.086} &
\cellcolor[RGB]{232,194,178}\makebox[2.8em][c]{0.087} &
\cellcolor[RGB]{255,255,255}\makebox[2.8em][c]{0.001} \\[2pt]
\cellcolor[RGB]{147,153,229}\makebox[2.8em][c]{-0.153} &
\cellcolor[RGB]{255,255,255}\makebox[2.8em][c]{0.000} &
\cellcolor[RGB]{255,255,255}\makebox[2.8em][c]{0.001} &
\cellcolor[RGB]{192,110,88}\makebox[2.8em][c]{0.503}
\end{array}\right]
+i\,
\left[\begin{array}{cccc}
\cellcolor[RGB]{255,255,255}\makebox[2.8em][c]{0} &
\cellcolor[RGB]{245,246,251}\makebox[2.8em][c]{-0.006} &
\cellcolor[RGB]{245,246,251}\makebox[2.8em][c]{-0.006} &
\cellcolor[RGB]{254,254,254}\makebox[2.8em][c]{-0.001} \\[2pt]
\cellcolor[RGB]{245,244,243}\makebox[2.8em][c]{0.006} &
\cellcolor[RGB]{255,255,255}\makebox[2.8em][c]{0} &
\cellcolor[RGB]{254,254,255}\makebox[2.8em][c]{-0.000} &
\cellcolor[RGB]{147,153,229}\makebox[2.8em][c]{-0.119} \\[2pt]
\cellcolor[RGB]{245,244,243}\makebox[2.8em][c]{0.006} &
\cellcolor[RGB]{255,255,255}\makebox[2.8em][c]{0.000} &
\cellcolor[RGB]{255,255,255}\makebox[2.8em][c]{0} &
\cellcolor[RGB]{147,153,229}\makebox[2.8em][c]{-0.118} \\[2pt]
\cellcolor[RGB]{254,254,254}\makebox[2.8em][c]{0.001} &
\cellcolor[RGB]{192,110,88}\makebox[2.8em][c]{0.119} &
\cellcolor[RGB]{192,110,88}\makebox[2.8em][c]{0.118} &
\cellcolor[RGB]{255,255,255}\makebox[2.8em][c]{0}
\end{array}\right]
\]

\[
C =
\begin{pmatrix}
0.135 & -0.001 & 0.002 \\
-0.002 & -0.479 & 0.227 \\
0.004 & 0.224 & -0.651
\end{pmatrix}
\pm 0.002 \quad
B^{+} =
\begin{pmatrix}
-0.001 \\
0.251 \\
-0.179
\end{pmatrix}
\pm 0.001 \quad
B^{-} =
\begin{pmatrix}
-0.002 \\
0.249 \\
-0.180
\end{pmatrix}
\pm 0.001
\]

\[
\begin{array}{l}
\boldsymbol{e^{+}e^{-} \rightarrow t\bar{t} \quad (\sqrt{s}=1~\textbf{TeV}):}
\end{array} \hfill
\rho_{t\bar{t}} =
\left[\begin{array}{cccc}
\cellcolor[RGB]{202,134,116}\makebox[2.8em][c]{0.357} &
\cellcolor[RGB]{255,255,255}\makebox[2.8em][c]{-0.000} &
\cellcolor[RGB]{254,254,255}\makebox[2.8em][c]{-0.002} &
\cellcolor[RGB]{149,154,230}\makebox[2.8em][c]{-0.147} \\[2pt]
\cellcolor[RGB]{255,255,255}\makebox[2.8em][c]{-0.000} &
\cellcolor[RGB]{251,241,240}\makebox[2.8em][c]{0.021} &
\cellcolor[RGB]{251,240,239}\makebox[2.8em][c]{0.023} &
\cellcolor[RGB]{254,254,254}\makebox[2.8em][c]{0.002} \\[2pt]
\cellcolor[RGB]{254,254,255}\makebox[2.8em][c]{-0.002} &
\cellcolor[RGB]{251,240,239}\makebox[2.8em][c]{0.023} &
\cellcolor[RGB]{251,241,240}\makebox[2.8em][c]{0.022} &
\cellcolor[RGB]{255,255,255}\makebox[2.8em][c]{-0.000} \\[2pt]
\cellcolor[RGB]{149,154,230}\makebox[2.8em][c]{-0.147} &
\cellcolor[RGB]{254,254,254}\makebox[2.8em][c]{0.002} &
\cellcolor[RGB]{255,255,255}\makebox[2.8em][c]{-0.000} &
\cellcolor[RGB]{192,110,88}\makebox[2.8em][c]{0.599}
\end{array}\right]
+i\,
\left[\begin{array}{cccc}
\cellcolor[RGB]{255,255,255}\makebox[2.8em][c]{0} &
\cellcolor[RGB]{245,244,243}\makebox[2.8em][c]{0.006} &
\cellcolor[RGB]{248,247,246}\makebox[2.8em][c]{0.004} &
\cellcolor[RGB]{255,255,255}\makebox[2.8em][c]{0.000} \\[2pt]
\cellcolor[RGB]{245,246,251}\makebox[2.8em][c]{-0.006} &
\cellcolor[RGB]{255,255,255}\makebox[2.8em][c]{0} &
\cellcolor[RGB]{254,254,254}\makebox[2.8em][c]{-0.001} &
\cellcolor[RGB]{143,166,233}\makebox[2.8em][c]{-0.069} \\[2pt]
\cellcolor[RGB]{248,247,246}\makebox[2.8em][c]{-0.004} &
\cellcolor[RGB]{255,255,255}\makebox[2.8em][c]{0.001} &
\cellcolor[RGB]{255,255,255}\makebox[2.8em][c]{0} &
\cellcolor[RGB]{141,164,233}\makebox[2.8em][c]{-0.070} \\[2pt]
\cellcolor[RGB]{255,255,255}\makebox[2.8em][c]{-0.000} &
\cellcolor[RGB]{236,172,154}\makebox[2.8em][c]{0.069}   &
\cellcolor[RGB]{235,170,152}\makebox[2.8em][c]{0.070}   &
\cellcolor[RGB]{255,255,255}\makebox[2.8em][c]{0}
\end{array}\right]
\]

\[
C =
\begin{pmatrix}
0.249 & 0.002 & 0.008 \\
-0.001 & -0.339 & 0.146 \\
-0.000 & 0.152 & -0.913
\end{pmatrix}
\pm 0.004 \quad
B^{+} =
\begin{pmatrix}
0.000 \\
0.129 \\
-0.243
\end{pmatrix}
\pm 0.002 \quad
B^{-} =
\begin{pmatrix}
-0.001 \\
0.126 \\
-0.241
\end{pmatrix}
\pm 0.002
\]

\vspace{1em}

\subsection*{\large Inclusive $\rho_{t\bar{t}}$ reconstruction for $\boldsymbol{e^{+}e^{-} \rightarrow t \bar{t} + g}$}

\[
\begin{array}{l}
\boldsymbol{e^{+}e^{-} \rightarrow t\bar{t}g \quad (\sqrt{s}=500~\textbf{GeV}):} \\
\quad p_T^g > 30~\text{GeV}
\end{array} \hfill
\rho_{t\bar{t}} =
\left[\begin{array}{cccc}
\cellcolor[RGB]{202,134,116}\makebox[2.8em][c]{0.301} &
\cellcolor[RGB]{254,254,255}\makebox[2.8em][c]{-0.000} &
\cellcolor[RGB]{255,255,255}\makebox[2.8em][c]{0.000} &
\cellcolor[RGB]{156,159,233}\makebox[2.8em][c]{-0.132} \\[2pt]
\cellcolor[RGB]{254,254,255}\makebox[2.8em][c]{-0.000} &
\cellcolor[RGB]{227,184,167}\makebox[2.8em][c]{0.113} &
\cellcolor[RGB]{246,228,223}\makebox[2.8em][c]{0.054} &
\cellcolor[RGB]{255,255,255}\makebox[2.8em][c]{0.000} \\[2pt]
\cellcolor[RGB]{255,255,255}\makebox[2.8em][c]{0.000} &
\cellcolor[RGB]{246,228,223}\makebox[2.8em][c]{0.054} &
\cellcolor[RGB]{227,185,168}\makebox[2.8em][c]{0.111} &
\cellcolor[RGB]{255,255,255}\makebox[2.8em][c]{0.001} \\[2pt]
\cellcolor[RGB]{156,159,233}\makebox[2.8em][c]{-0.132} &
\cellcolor[RGB]{255,255,255}\makebox[2.8em][c]{0.000} &
\cellcolor[RGB]{255,255,255}\makebox[2.8em][c]{0.001} &
\cellcolor[RGB]{197,120,99}\makebox[2.8em][c]{0.475}
\end{array}\right]
+i\,
\left[\begin{array}{cccc}
\cellcolor[RGB]{255,255,255}\makebox[2.8em][c]{0} &
\cellcolor[RGB]{250,250,253}\makebox[2.8em][c]{-0.002} &
\cellcolor[RGB]{250,250,253}\makebox[2.8em][c]{-0.003} &
\cellcolor[RGB]{254,254,254}\makebox[2.8em][c]{-0.000} \\[2pt]
\cellcolor[RGB]{250,249,249}\makebox[2.8em][c]{0.002} &
\cellcolor[RGB]{255,255,255}\makebox[2.8em][c]{0} &
\cellcolor[RGB]{254,254,254}\makebox[2.8em][c]{0.000} &
\cellcolor[RGB]{147,153,229}\makebox[2.8em][c]{-0.109} \\[2pt]
\cellcolor[RGB]{250,249,249}\makebox[2.8em][c]{0.003} &
\cellcolor[RGB]{254,254,254}\makebox[2.8em][c]{-0.000} &
\cellcolor[RGB]{255,255,255}\makebox[2.8em][c]{0} &
\cellcolor[RGB]{147,153,229}\makebox[2.8em][c]{-0.109} \\[2pt]
\cellcolor[RGB]{254,254,254}\makebox[2.8em][c]{0.000} &
\cellcolor[RGB]{192,110,88}\makebox[2.8em][c]{0.109} &
\cellcolor[RGB]{192,110,88}\makebox[2.8em][c]{0.109} &
\cellcolor[RGB]{255,255,255}\makebox[2.8em][c]{0}
\end{array}\right]
\]

\[
C =
\begin{pmatrix}
0.158 & -0.002 & -0.000 \\
0.000 & -0.372 & 0.212 \\
0.002 & 0.213 & -0.553
\end{pmatrix}
\pm 0.01 \quad
B^{+} =
\begin{pmatrix}
0.001 \\
0.223 \\
-0.173
\end{pmatrix}
\pm 0.007 \quad
B^{-} =
\begin{pmatrix}
0.000 \\
0.222 \\
-0.176
\end{pmatrix}
\pm 0.007
\]

\vspace{1em}

\[
\begin{array}{l}
\boldsymbol{e^{+}e^{-} \rightarrow t\bar{t}g \quad (\sqrt{s}=1~\textbf{TeV}):} \\
\quad p_T^g > 120~\text{GeV}
\end{array} \hfill
\rho_{t\bar{t}} =
\left[\begin{array}{cccc}
\cellcolor[RGB]{188,117,95}\makebox[2.8em][c]{0.316} &
\cellcolor[RGB]{255,255,255}\makebox[2.8em][c]{0.000} &
\cellcolor[RGB]{255,255,255}\makebox[2.8em][c]{-0.000} &
\cellcolor[RGB]{169,171,238}\makebox[2.8em][c]{-0.113} \\[2pt]
\cellcolor[RGB]{255,255,255}\makebox[2.8em][c]{0.000} &
\cellcolor[RGB]{232,194,178}\makebox[2.8em][c]{0.072} &
\cellcolor[RGB]{244,245,250}\makebox[2.8em][c]{-0.007} &
\cellcolor[RGB]{255,255,255}\makebox[2.8em][c]{-0.000} \\[2pt]
\cellcolor[RGB]{255,255,255}\makebox[2.8em][c]{-0.000} &
\cellcolor[RGB]{244,245,250}\makebox[2.8em][c]{-0.007} &
\cellcolor[RGB]{231,193,177}\makebox[2.8em][c]{0.074} &
\cellcolor[RGB]{254,254,255}\makebox[2.8em][c]{-0.001} \\[2pt]
\cellcolor[RGB]{169,171,238}\makebox[2.8em][c]{-0.113} &
\cellcolor[RGB]{255,255,255}\makebox[2.8em][c]{-0.000} &
\cellcolor[RGB]{254,254,255}\makebox[2.8em][c]{-0.001} &
\cellcolor[RGB]{192,110,88}\makebox[2.8em][c]{0.538}
\end{array}\right]
+i\,
\left[\begin{array}{cccc}
\cellcolor[RGB]{255,255,255}\makebox[2.8em][c]{0} &
\cellcolor[RGB]{245,244,243}\makebox[2.8em][c]{0.007} &
\cellcolor[RGB]{245,244,243}\makebox[2.8em][c]{0.008} &
\cellcolor[RGB]{252,251,251}\makebox[2.8em][c]{0.002} \\[2pt]
\cellcolor[RGB]{244,245,250}\makebox[2.8em][c]{-0.007} &
\cellcolor[RGB]{255,255,255}\makebox[2.8em][c]{0} &
\cellcolor[RGB]{252,251,251}\makebox[2.8em][c]{0.002} &
\cellcolor[RGB]{143,165,233}\makebox[2.8em][c]{-0.062} \\[2pt]
\cellcolor[RGB]{242,243,249}\makebox[2.8em][c]{-0.008} &
\cellcolor[RGB]{252,251,251}\makebox[2.8em][c]{-0.002} &
\cellcolor[RGB]{255,255,255}\makebox[2.8em][c]{0} &
\cellcolor[RGB]{141,164,233}\makebox[2.8em][c]{-0.063} \\[2pt]
\cellcolor[RGB]{252,251,251}\makebox[2.8em][c]{-0.002} &
\cellcolor[RGB]{236,172,154}\makebox[2.8em][c]{0.062} &
\cellcolor[RGB]{236,172,154}\makebox[2.8em][c]{0.063} &
\cellcolor[RGB]{255,255,255}\makebox[2.8em][c]{0}
\end{array}\right]
\]

\[
C =
\begin{pmatrix}
0.24 & 0.00 & -0.00 \\
0.01 & -0.21 & 0.14 \\
-0.00 & 0.14 & -0.71
\end{pmatrix}
\pm 0.02 \quad
B^{+} =
\begin{pmatrix}
-0.001 \\
0.107 \\
-0.223
\end{pmatrix}
\pm 0.009 \quad
B^{-} =
\begin{pmatrix}
-0.002 \\
0.113 \\
-0.220
\end{pmatrix}
\pm 0.009
\]

\vspace{1em}

\subsection*{\large Numerical results for decoherence between $\boldsymbol{e^{+}e^{-} \rightarrow t \bar{t}}$ and $\boldsymbol{e^{+}e^{-} \rightarrow t \bar{t} + g}$ using the $D_n$ marker.}

\begingroup
\def\arraystretch{1.3}

\begin{table*}[!h] \leftskip-.5em
\begin{tabular}{|c|c|cc|c|c|cc|cc|cc|cc|c|}
\hline
Machine & $\sqrt{s}$ & \multicolumn{2}{c|}{$\sigma   (pb) \footnote{A selection }$ } & $\int \mathcal{L}dt$ & \multirow{2}{*}{Decay channel   \footnote{Naming follows from the used polarimeter ($\alpha = 1$): $l \equiv   e, \mu ;~ q_d \equiv d, s$. A 60\% acceptance is considered for   the hadronic polarimeters.}} & \multicolumn{2}{c|}{events   ($10^6$)} & \multicolumn{1}{c|}{$D_n$} & $\Delta D_n$ & \multicolumn{1}{c|}{$D_n$} & $\Delta D_n$ & \multicolumn{2}{c|}{$D_n^{t\bar{t}g}/D_n^{t\bar{t}}$} & $\mathcal{S}$ \\ \cline{3-4} \cline{7-14}
linear $e^{+}e^{-}$ & [TeV] & \multicolumn{1}{c|}{$t\bar{t}$} & $t\bar{t}g$ & [ab$^{-1}$] &  & \multicolumn{1}{c|}{$t\bar{t}$} & $t\bar{t}g$ & \multicolumn{2}{c|}{$t\bar{t}$} & \multicolumn{2}{c|}{$t\bar{t}g$} & \multicolumn{1}{c|}{$R_{D_n}$} & $\Delta R_{D_n}$ &  \\ \hline
\multirow{3}{*}{LCF~\cite{LinearColliderVision:2025hlt}} & \multirow{3}{*}{0.5} & \multicolumn{1}{c|}{\multirow{3}{*}{$0.5482$}} & \multirow{3}{*}{$0.0174$} & \multirow{3}{*}{8} & $l^+l^-$ & \multicolumn{1}{c|}{$0.22$} & $0.01$ & \multicolumn{1}{c|}{\multirow{3}{*}{$-0.422$}} & $0.003$ & \multicolumn{1}{c|}{\multirow{3}{*}{$-0.360$ \footnote{$p_T^g > 30$ GeV}}} & $0.019 $ & \multicolumn{1}{c|}{\multirow{3}{*}{$0.86$}} & $0.05$ & $3.0\,\sigma$ \\ \cline{6-8} \cline{10-10} \cline{12-12} \cline{14-15} 
 &  & \multicolumn{1}{c|}{} &  &  & $l^+ l^-, l q_d$ & \multicolumn{1}{c|}{$1.10$} & $0.03$ & \multicolumn{1}{c|}{} & $0.002$ & \multicolumn{1}{c|}{} & $0.009$ & \multicolumn{1}{c|}{} & $0.02$ & $6.4\,\sigma$ \\ \cline{6-8} \cline{10-10} \cline{12-12} \cline{14-15} 
 &  & \multicolumn{1}{c|}{} &  &  & $l^+ l^-, l q_d, q_d q_d$ & \multicolumn{1}{c|}{$1.71$} & $0.05$ & \multicolumn{1}{c|}{} & $0.002$ & \multicolumn{1}{c|}{} & $0.007$ & \multicolumn{1}{c|}{} & $0.02$ & $8.3\,\sigma$ \\ \hline
\multirow{3}{*}{\begin{tabular}[c]{@{}c@{}}LCF\\      (Upgrade) ~\cite{LinearColliderVision:2025hlt}\end{tabular}} & \multirow{3}{*}{1} & \multicolumn{1}{c|}{\multirow{3}{*}{$0.1654$}} & \multirow{3}{*}{$0.0121$} & \multirow{3}{*}{8} & $l^+l^-$ & \multicolumn{1}{c|}{$0.07$} & $<0.01$ & \multicolumn{1}{c|}{\multirow{3}{*}{$-0.500$}} & $0.006$ & \multicolumn{1}{c|}{\multirow{3}{*}{$-0.39$ \footnote{$p_T^g > 120$ GeV}}} & $0.030$ & \multicolumn{1}{c|}{\multirow{3}{*}{$0.78$}} & $0.05$ & $4.6\,\sigma$ \\ \cline{6-8} \cline{10-10} \cline{12-12} \cline{14-15} 
 &  & \multicolumn{1}{c|}{} &  &  & $l^+ l^-, l q_d$ & \multicolumn{1}{c|}{$0.30$} & $0.02$ & \multicolumn{1}{c|}{} & $0.003$ & \multicolumn{1}{c|}{} & $0.011$ & \multicolumn{1}{c|}{} & $0.02$ & $9.8\,\sigma$ \\ \cline{6-8} \cline{10-10} \cline{12-12} \cline{14-15} 
 &  & \multicolumn{1}{c|}{} &  &  & $l^+ l^-, l q_d, q_d q_d$ & \multicolumn{1}{c|}{$0.52$} & $0.04$ & \multicolumn{1}{c|}{} & $0.002$ & \multicolumn{1}{c|}{} & $0.009$ & \multicolumn{1}{c|}{} & $0.02$ & $12.7\,\sigma$ \\ \hline
\end{tabular}
\caption{Projected statistical significance for the loss of entanglement between $t\bar{t}$ and $t\bar{t} + g$  production in $e^+ e^- $ collisions considering the di-leptonic, semilecptonic and inclusive final states, assuming a 60 \% down-type quark acceptance. $D_n$ marker predictions are provided for the inclusive $t \bar{t}$ sample and with an optimized $p_T$ cut for $t\bar{t} + g$.}
\end{table*}
\endgroup

\subsection*{\large Values for $\boldsymbol{p p \rightarrow t \bar{t}}$}

\[
\begin{array}{l}
\boldsymbol{pp \rightarrow t\bar{t} \quad (\sqrt{s}=13~\textbf{TeV}):} \\
\text{Fiducial region:} \\ 
\quad m_{t\bar{t}} > 800~\text{GeV} \\
\quad \left|\cos\!\left(\Theta_t^{t\bar{t}\mathrm{RF}}\right)\right| \leq 0.4
\end{array} \hfill
\rho_{t\bar{t}} =
\left[\begin{array}{cccc}
\cellcolor[RGB]{209,149,133}\makebox[2.8em][c]{0.365} &
\cellcolor[RGB]{249,250,253}\makebox[2.8em][c]{-0.005} &
\cellcolor[RGB]{254,253,253}\makebox[2.8em][c]{0.006} &
\cellcolor[RGB]{60,78,194}\color{white}\makebox[2.8em][c]{-0.332} \\[2pt]
\cellcolor[RGB]{249,250,253}\makebox[2.8em][c]{-0.005} &
\cellcolor[RGB]{239,218,213}\makebox[2.8em][c]{0.127} &
\cellcolor[RGB]{251,245,244}\makebox[2.8em][c]{0.034} &
\cellcolor[RGB]{255,254,254}\makebox[2.8em][c]{0.003} \\[2pt]
\cellcolor[RGB]{254,253,253}\makebox[2.8em][c]{0.006} &
\cellcolor[RGB]{251,245,244}\makebox[2.8em][c]{0.034} &
\cellcolor[RGB]{238,216,211}\makebox[2.8em][c]{0.133} &
\cellcolor[RGB]{250,250,253}\makebox[2.8em][c]{-0.004} \\[2pt]
\cellcolor[RGB]{60,78,194}\color{white}\makebox[2.8em][c]{-0.332} &
\cellcolor[RGB]{255,254,254}\makebox[2.8em][c]{0.003} &
\cellcolor[RGB]{250,250,253}\makebox[2.8em][c]{-0.004} &
\cellcolor[RGB]{208,146,130}\makebox[2.8em][c]{0.375}
\end{array}\right]
+i\,
\left[\begin{array}{cccc}
\cellcolor[RGB]{255,255,255}\makebox[2.8em][c]{0} &
\cellcolor[RGB]{198,203,237}\makebox[2.8em][c]{-0.035} &
\cellcolor[RGB]{221,224,244}\makebox[2.8em][c]{-0.021} &
\cellcolor[RGB]{247,247,252}\makebox[2.8em][c]{-0.005} \\[2pt]
\cellcolor[RGB]{236,212,206}\makebox[2.8em][c]{0.035} &
\cellcolor[RGB]{255,255,255}\makebox[2.8em][c]{0} &
\cellcolor[RGB]{250,250,253}\makebox[2.8em][c]{-0.003} &
\cellcolor[RGB]{242,225,220}\makebox[2.8em][c]{0.025} \\[2pt]
\cellcolor[RGB]{244,230,226}\makebox[2.8em][c]{0.021} &
\cellcolor[RGB]{253,251,251}\makebox[2.8em][c]{0.003} &
\cellcolor[RGB]{255,255,255}\makebox[2.8em][c]{0} &
\cellcolor[RGB]{237,213,207}\makebox[2.8em][c]{0.035} \\[2pt]
\cellcolor[RGB]{252,249,248}\makebox[2.8em][c]{0.005} &
\cellcolor[RGB]{215,218,242}\makebox[2.8em][c]{-0.025} &
\cellcolor[RGB]{198,204,237}\makebox[2.8em][c]{-0.035} &
\cellcolor[RGB]{255,255,255}\makebox[2.8em][c]{0}
\end{array}\right]
\]

\[
C =
\begin{pmatrix}
0.596 & -0.004 & -0.006 \\
-0.017 & -0.733 & -0.092 \\
0.001 & -0.140 & -0.479
\end{pmatrix}
\pm 0.006 \quad
B^{+} =
\begin{pmatrix}
0.019 \\
-0.008 \\
-0.016
\end{pmatrix}
\pm 0.004 \quad
B^{-} =
\begin{pmatrix}
-0.017 \\
0.001 \\
-0.004
\end{pmatrix}
\pm 0.004
\]

\[
\begin{array}{l}
\boldsymbol{pp \rightarrow t\bar{t} \quad (\sqrt{s}=14~\textbf{TeV}):} \\
\text{Fiducial region:} \\ 
\quad m_{t\bar{t}} > 800~\text{GeV} \\
\quad \left|\cos\!\left(\Theta_t^{t\bar{t}\mathrm{RF}}\right)\right| \leq 0.4
\end{array} \hfill
\rho_{t\bar{t}} =
\left[\begin{array}{cccc}
\cellcolor[RGB]{208,146,130}\makebox[2.8em][c]{0.379} &
\cellcolor[RGB]{245,246,250}\makebox[2.8em][c]{-0.008} &
\cellcolor[RGB]{255,254,254}\makebox[2.8em][c]{0.004} &
\cellcolor[RGB]{60,78,194}\color{white}\makebox[2.8em][c]{-0.336} \\[2pt]
\cellcolor[RGB]{245,246,250}\makebox[2.8em][c]{-0.008} &
\cellcolor[RGB]{240,219,214}\makebox[2.8em][c]{0.119} &
\cellcolor[RGB]{252,247,246}\makebox[2.8em][c]{0.028} &
\cellcolor[RGB]{246,247,251}\makebox[2.8em][c]{-0.008} \\[2pt]
\cellcolor[RGB]{255,254,254}\makebox[2.8em][c]{0.004} &
\cellcolor[RGB]{252,247,246}\makebox[2.8em][c]{0.028} &
\cellcolor[RGB]{240,218,213}\makebox[2.8em][c]{0.123} &
\cellcolor[RGB]{255,254,254}\makebox[2.8em][c]{0.003} \\[2pt]
\cellcolor[RGB]{60,78,194}\color{white}\makebox[2.8em][c]{-0.336} &
\cellcolor[RGB]{246,247,251}\makebox[2.8em][c]{-0.008} &
\cellcolor[RGB]{255,254,254}\makebox[2.8em][c]{0.003} &
\cellcolor[RGB]{208,146,130}\makebox[2.8em][c]{0.380}
\end{array}\right]
+i\,
\left[\begin{array}{cccc}
\cellcolor[RGB]{255,255,255}\makebox[2.8em][c]{0} &
\cellcolor[RGB]{215,218,242}\makebox[2.8em][c]{-0.025} &
\cellcolor[RGB]{202,207,239}\makebox[2.8em][c]{-0.032} &
\cellcolor[RGB]{249,249,251}\makebox[2.8em][c]{0.011} \\[2pt]
\cellcolor[RGB]{242,225,220}\makebox[2.8em][c]{0.025} &
\cellcolor[RGB]{255,255,255}\makebox[2.8em][c]{0} &
\cellcolor[RGB]{253,252,252}\makebox[2.8em][c]{-0.003} &
\cellcolor[RGB]{234,206,199}\makebox[2.8em][c]{0.038} \\[2pt]
\cellcolor[RGB]{236,212,206}\makebox[2.8em][c]{0.032} &
\cellcolor[RGB]{253,252,252}\makebox[2.8em][c]{0.003} &
\cellcolor[RGB]{255,255,255}\makebox[2.8em][c]{0} &
\cellcolor[RGB]{242,224,219}\makebox[2.8em][c]{0.025} \\[2pt]
\cellcolor[RGB]{249,249,251}\makebox[2.8em][c]{-0.011} &
\cellcolor[RGB]{194,200,236}\makebox[2.8em][c]{-0.038} &
\cellcolor[RGB]{214,218,242}\makebox[2.8em][c]{-0.025} &
\cellcolor[RGB]{255,255,255}\makebox[2.8em][c]{0}
\end{array}\right]
\]

\[
C =
\begin{pmatrix}
0.617 & 0.027 & -0.024 \\
0.016 & -0.727 & -0.141 \\
0.023 & -0.100 & -0.517
\end{pmatrix}
\pm 0.001 \quad
B^{+} =
\begin{pmatrix}
-0.0064 \\
-0.0115 \\
-0.0046
\end{pmatrix}
\pm 0.0007 \quad
B^{-} =
\begin{pmatrix}
-0.0101 \\
-0.0012 \\
0.0025
\end{pmatrix}
\pm 0.0007
\]

\vspace{1em}

\subsection*{\large Values for $\boldsymbol{p p \rightarrow t \bar{t}j}$}

\[
\begin{array}{l}
\boldsymbol{pp \rightarrow t\bar{t}j \quad (\sqrt{s}=14~\textbf{TeV}):} \\
\text{Fiducial region:} \\
\quad m_{t\bar{t}} > 800~\text{GeV} \\
\quad \left|\cos\!\left(\Theta_t^{t\bar{t}\mathrm{RF}}\right)\right| \leq 0.4 \\
p_T^j > 250~\text{GeV}
\end{array} \hfill
\rho_{t\bar{t}} =
\left[\begin{array}{cccc}
\cellcolor[RGB]{210,151,136}\makebox[2.8em][c]{0.360} &
\cellcolor[RGB]{255,255,255}\makebox[2.8em][c]{0.001} &
\cellcolor[RGB]{255,254,254}\makebox[2.8em][c]{0.005} &
\cellcolor[RGB]{115,127,217}\makebox[2.8em][c]{-0.195} \\[2pt]
\cellcolor[RGB]{255,255,255}\makebox[2.8em][c]{0.001} &
\cellcolor[RGB]{238,215,210}\makebox[2.8em][c]{0.135} &
\cellcolor[RGB]{251,246,245}\makebox[2.8em][c]{0.031} &
\cellcolor[RGB]{255,254,254}\makebox[2.8em][c]{0.003} \\[2pt]
\cellcolor[RGB]{255,254,254}\makebox[2.8em][c]{0.005} &
\cellcolor[RGB]{251,246,245}\makebox[2.8em][c]{0.031} &
\cellcolor[RGB]{237,214,208}\makebox[2.8em][c]{0.138} &
\cellcolor[RGB]{255,254,254}\makebox[2.8em][c]{0.003} \\[2pt]
\cellcolor[RGB]{115,127,217}\makebox[2.8em][c]{-0.195} &
\cellcolor[RGB]{255,254,254}\makebox[2.8em][c]{0.003} &
\cellcolor[RGB]{255,254,254}\makebox[2.8em][c]{0.003} &
\cellcolor[RGB]{209,149,133}\makebox[2.8em][c]{0.366}
\end{array}\right]
+i\,
\left[\begin{array}{cccc}
\cellcolor[RGB]{255,255,255}\makebox[2.8em][c]{0} &
\cellcolor[RGB]{248,248,252}\makebox[2.8em][c]{-0.004} &
\cellcolor[RGB]{236,239,248}\makebox[2.8em][c]{-0.014} &
\cellcolor[RGB]{250,249,248}\makebox[2.8em][c]{0.010} \\[2pt]
\cellcolor[RGB]{249,248,247}\makebox[2.8em][c]{0.004} &
\cellcolor[RGB]{255,255,255}\makebox[2.8em][c]{0} &
\cellcolor[RGB]{248,250,252}\makebox[2.8em][c]{0.009} &
\cellcolor[RGB]{244,229,225}\makebox[2.8em][c]{0.016} \\[2pt]
\cellcolor[RGB]{243,230,226}\makebox[2.8em][c]{0.014} &
\cellcolor[RGB]{248,250,252}\makebox[2.8em][c]{-0.009} &
\cellcolor[RGB]{255,255,255}\makebox[2.8em][c]{0} &
\cellcolor[RGB]{249,249,252}\makebox[2.8em][c]{-0.004} \\[2pt]
\cellcolor[RGB]{250,249,248}\makebox[2.8em][c]{-0.010} &
\cellcolor[RGB]{235,238,248}\makebox[2.8em][c]{-0.016} &
\cellcolor[RGB]{249,248,247}\makebox[2.8em][c]{0.004} &
\cellcolor[RGB]{255,255,255}\makebox[2.8em][c]{0}
\end{array}\right]
\]

\[
C =
\begin{pmatrix}
0.33 & 0.00 & -0.00 \\
0.04 & -0.45 & -0.06 \\
0.00 & -0.00 & -0.45
\end{pmatrix}
\pm 0.02 \quad
B^{+} =
\begin{pmatrix}
0.01 \\
-0.00 \\
-0.01
\end{pmatrix}
\pm 0.01 \quad
B^{-} =
\begin{pmatrix}
0.01 \\
0.02 \\
-0.00
\end{pmatrix}
\pm 0.01
\]

\[
\begin{array}{l}
\boldsymbol{pp \rightarrow t\bar{t}j \quad (\sqrt{s}=14~\textbf{TeV}):} \\
\text{Fiducial region:} \\
\quad m_{t\bar{t}} > 800~\text{GeV} \\
\quad \left|\cos\!\left(\Theta_t^{t\bar{t}\mathrm{RF}}\right)\right| \leq 0.4 \\
p_T^j > 250~\text{GeV}
\end{array} \hfill
\rho_{t\bar{t}} =
\left[\begin{array}{cccc}
\cellcolor[RGB]{205,140,123}\makebox[2.8em][c]{0.393} &
\cellcolor[RGB]{244,245,250}\makebox[2.8em][c]{-0.009} &
\cellcolor[RGB]{253,252,251}\makebox[2.8em][c]{0.011} &
\cellcolor[RGB]{147,153,229}\makebox[2.8em][c]{-0.154} \\[2pt]
\cellcolor[RGB]{244,245,250}\makebox[2.8em][c]{-0.009} &
\cellcolor[RGB]{239,218,213}\makebox[2.8em][c]{0.125} &
\cellcolor[RGB]{235,239,248}\makebox[2.8em][c]{-0.049} &
\cellcolor[RGB]{254,253,253}\makebox[2.8em][c]{0.007} \\[2pt]
\cellcolor[RGB]{253,252,251}\makebox[2.8em][c]{0.011} &
\cellcolor[RGB]{235,239,248}\makebox[2.8em][c]{-0.049} &
\cellcolor[RGB]{239,217,212}\makebox[2.8em][c]{0.126} &
\cellcolor[RGB]{242,243,249}\makebox[2.8em][c]{-0.042} \\[2pt]
\cellcolor[RGB]{147,153,229}\makebox[2.8em][c]{-0.154} &
\cellcolor[RGB]{254,253,253}\makebox[2.8em][c]{0.007} &
\cellcolor[RGB]{242,243,249}\makebox[2.8em][c]{-0.042} &
\cellcolor[RGB]{210,152,137}\makebox[2.8em][c]{0.356}
\end{array}\right]
+i\,
\left[\begin{array}{cccc}
\cellcolor[RGB]{255,255,255}\makebox[2.8em][c]{0} &
\cellcolor[RGB]{234,209,202}\makebox[2.8em][c]{0.033} &
\cellcolor[RGB]{191,198,235}\makebox[2.8em][c]{-0.040} &
\cellcolor[RGB]{227,198,190}\makebox[2.8em][c]{0.053} \\[2pt]
\cellcolor[RGB]{211,215,241}\makebox[2.8em][c]{-0.033} &
\cellcolor[RGB]{255,255,255}\makebox[2.8em][c]{0} &
\cellcolor[RGB]{237,213,207}\makebox[2.8em][c]{0.023} &
\cellcolor[RGB]{232,203,195}\makebox[2.8em][c]{0.048} \\[2pt]
\cellcolor[RGB]{223,194,186}\makebox[2.8em][c]{0.040} &
\cellcolor[RGB]{218,221,243}\makebox[2.8em][c]{-0.023} &
\cellcolor[RGB]{255,255,255}\makebox[2.8em][c]{0} &
\cellcolor[RGB]{234,209,202}\makebox[2.8em][c]{0.027} \\[2pt]
\cellcolor[RGB]{182,190,232}\makebox[2.8em][c]{-0.053} &
\cellcolor[RGB]{177,186,231}\makebox[2.8em][c]{-0.048} &
\cellcolor[RGB]{211,215,241}\makebox[2.8em][c]{-0.027} &
\cellcolor[RGB]{255,255,255}\makebox[2.8em][c]{0}
\end{array}\right]
\]

\[
C =
\begin{pmatrix}
0.406 & 0.060 & -0.010 \\
0.152 & -0.210 & -0.176 \\
-0.066 & 0.013 & -0.499
\end{pmatrix}
\pm 0.004 \quad
B^{+} =
\begin{pmatrix}
0.036 \\
-0.015 \\
0.036
\end{pmatrix}
\pm 0.002 \quad
B^{-} =
\begin{pmatrix}
-0.103 \\
-0.121 \\
0.038
\end{pmatrix}
\pm 0.002
\]

\subsection*{\large Numerical results for decoherence between $\boldsymbol{p p \rightarrow t \bar{t}}$ and $\boldsymbol{p p \rightarrow t \bar{t}j}$ using the $D_n$ marker.}

\begingroup
\def\arraystretch{1.25}
\begin{table*}[!h] \leftskip-0em
\begin{tabular}{|c|c|cc|c|c|cc|cc|cc|cc|c|}
\hline
Machine & $\sqrt{s}$ & \multicolumn{2}{c|}{$\sigma   (pb) \footnote{Fiducial region:   $|\cos\Theta_{t}^{t\bar{t}RF}|<0.4, m_{t\bar{t}} > 800$ GeV} $} & $\int \mathcal{L}dt$ & \multirow{2}{*}{Decay channel \footnote{Naming follows from the used polarimeter ($\alpha =   1$): $l \equiv e, \mu ;~ q_d \equiv d, s$. A 60\% acceptance is considered for each hadronic polarimeters.}} & \multicolumn{2}{c|}{events   ($10^6$)} & \multicolumn{1}{c|}{$D_n$} & $\Delta D_n$ & \multicolumn{1}{c|}{$D_n$} & $\Delta D_n$ & \multicolumn{2}{c|}{$R_{D_n} = D_n^{t\bar{t}j}/D_n^{t\bar{t}}$} & \multirow{2}{*}{$\mathcal{S}$} \\ \cline{1-1} \cline{3-4} \cline{7-14}
LHC & [TeV] & \multicolumn{1}{c|}{$t\bar{t}$} & $t\bar{t}j$ & [ab$^{-1}$] &  & \multicolumn{1}{c|}{$t\bar{t}$} & $t\bar{t}j$ & \multicolumn{2}{c|}{$t\bar{t}$} & \multicolumn{2}{c|}{$t\bar{t}g$} & \multicolumn{1}{c|}{$R_{D_n}$} & $\Delta R_{D_n}$ &  \\ \hline
\multirow{3}{*}{LHC Run 2} & \multirow{3}{*}{13} & \multicolumn{1}{c|}{\multirow{3}{*}{$4.06$}} & \multirow{3}{*}{$0.40$} & \multirow{3}{*}{0.14} & $l^+l^-$ & \multicolumn{1}{c|}{$0.03$} & $<0.01$ & \multicolumn{1}{c|}{\multirow{3}{*}{$-0.602$}} & $0.009$ & \multicolumn{1}{c|}{\multirow{3}{*}{$-0.41$   \footnote{$p_T^j > 250$ GeV \label{fn:pp_pTcut}}}} & $0.03$ & \multicolumn{1}{c|}{\multirow{3}{*}{$0.68$}} & $0.05$ & $ 4.2\,\sigma$ \\ \cline{6-8} \cline{10-10} \cline{12-12} \cline{14-15} 
 &  & \multicolumn{1}{c|}{} &  &  & $l^+ l^-, l q_d$ & \multicolumn{1}{c|}{$0.13$} & $0.01$ & \multicolumn{1}{c|}{} & $0.004$ & \multicolumn{1}{c|}{} & $0.02$ & \multicolumn{1}{c|}{} & $0.03$ & $ 6.5\,\sigma$ \\ \cline{6-8} \cline{10-10} \cline{12-12} \cline{14-15} 
 &  & \multicolumn{1}{c|}{} &  &  & $l^+ l^-, l q_d, q_d q_d$ & \multicolumn{1}{c|}{$0.22$} & $0.02$ & \multicolumn{1}{c|}{} & $0.003$ & \multicolumn{1}{c|}{} & $0.01$ & \multicolumn{1}{c|}{} & $0.02$ & $ 13\,\sigma$ \\ \hline
\multirow{3}{*}{HL-LHC} & \multirow{3}{*}{14} & \multicolumn{1}{c|}{\multirow{3}{*}{$4.97$}} & \multirow{3}{*}{$0.59$} & \multirow{3}{*}{3} & $l^+l^-$ & \multicolumn{1}{c|}{$0.75$} & $0.09$ & \multicolumn{1}{c|}{\multirow{3}{*}{$-0.620$}} & $0.002$ & \multicolumn{1}{c|}{\multirow{3}{*}{$-0.372 ^{\text{\ref{fn:pp_pTcut}}}$}} & $0.006$ & \multicolumn{1}{c|}{\multirow{3}{*}{$0.60$}} & $0.01$ & $ 24\,\sigma$ \\ \cline{6-8} \cline{10-10} \cline{12-12} \cline{14-15} 
 &  & \multicolumn{1}{c|}{} &  &  & $l^+ l^-, l q_d$ & \multicolumn{1}{c|}{$3.44$} & $0.41$ & \multicolumn{1}{c|}{} & $0.0001$ & \multicolumn{1}{c|}{} & $0.003$ & \multicolumn{1}{c|}{} & $0.005$ & $ 48\,\sigma$ \\ \cline{6-8} \cline{10-10} \cline{12-12} \cline{14-15} 
 &  & \multicolumn{1}{c|}{} &  &  & $l^+ l^-, l q_d, q_d q_d$ & \multicolumn{1}{c|}{$5.81$} & $0.70$ & \multicolumn{1}{c|}{} & $0.0001$ & \multicolumn{1}{c|}{} & $0.002$ & \multicolumn{1}{c|}{} & $0.003$ & $ 70\,\sigma$ \\ \hline
\end{tabular}
\caption{Prospected statistical significance for the loss of entanglement between $t\bar{t}$ and $t\bar{t} + j$ production from pp collision, considering the dileptonic, semileptonic and inclusive final states, assuming a 60\% down-type quark acceptance. $D_n$ marker predictions are provided for the inclusive $t\bar{t}$ sample and with an optimized $p_T^{jet}$ cut for $t\bar{t} + j$. }
\end{table*}
\endgroup

\subsection*{\large Values for $\boldsymbol{e^{+}e^{-} \rightarrow \tau^{+} \tau^{-}}$}

\[
\begin{array}{l}
\boldsymbol{e^{+}e^{-} \rightarrow \tau^{+}\tau^{-}} \textbf{(Belle II)}\\ 
(\sqrt{s}\approx10.6~\text{GeV}): \\
\left|\cos\!\left(\Theta_{\tau}^{\tau^{+}\tau^{-}\mathrm{RF}}\right)\right| \leq 0.4
\end{array} \hfill
\rho_{\tau^{+}\tau^{-}} =
\left[\begin{array}{cccc}
\cellcolor[RGB]{196,121,100}\makebox[2.8em][c]{0.457} &
\cellcolor[RGB]{254,254,255}\makebox[2.8em][c]{-0.002} &
\cellcolor[RGB]{255,255,255}\makebox[2.8em][c]{0.001} &
\cellcolor[RGB]{60,78,194}\color{white}\makebox[2.8em][c]{-0.410} \\[2pt]
\cellcolor[RGB]{254,254,255}\makebox[2.8em][c]{-0.002} &
\cellcolor[RGB]{248,233,229}\makebox[2.8em][c]{0.047} &
\cellcolor[RGB]{248,234,229}\makebox[2.8em][c]{0.046} &
\cellcolor[RGB]{255,255,255}\makebox[2.8em][c]{0.002} \\[2pt]
\cellcolor[RGB]{255,255,255}\makebox[2.8em][c]{0.001} &
\cellcolor[RGB]{248,234,229}\makebox[2.8em][c]{0.046} &
\cellcolor[RGB]{248,234,230}\makebox[2.8em][c]{0.046} &
\cellcolor[RGB]{255,254,254}\makebox[2.8em][c]{0.001} \\[2pt]
\cellcolor[RGB]{60,78,194}\color{white}\makebox[2.8em][c]{-0.410} &
\cellcolor[RGB]{255,255,255}\makebox[2.8em][c]{0.002} &
\cellcolor[RGB]{255,254,254}\makebox[2.8em][c]{0.001} &
\cellcolor[RGB]{198,124,104}\makebox[2.8em][c]{0.451}
\end{array}\right]
+i\,
\left[\begin{array}{cccc}
\cellcolor[RGB]{255,255,255}\makebox[2.8em][c]{0} &
\cellcolor[RGB]{253,252,252}\makebox[2.8em][c]{-0.001} &
\cellcolor[RGB]{254,253,253}\makebox[2.8em][c]{0.001} &
\cellcolor[RGB]{254,254,254}\makebox[2.8em][c]{-0.000} \\[2pt]
\cellcolor[RGB]{253,253,253}\makebox[2.8em][c]{0.001} &
\cellcolor[RGB]{255,255,255}\makebox[2.8em][c]{0} &
\cellcolor[RGB]{254,254,254}\makebox[2.8em][c]{0.001} &
\cellcolor[RGB]{254,253,253}\makebox[2.8em][c]{0.001} \\[2pt]
\cellcolor[RGB]{254,254,254}\makebox[2.8em][c]{-0.001} &
\cellcolor[RGB]{254,254,254}\makebox[2.8em][c]{-0.001} &
\cellcolor[RGB]{255,255,255}\makebox[2.8em][c]{0} &
\cellcolor[RGB]{251,249,248}\makebox[2.8em][c]{0.002} \\[2pt]
\cellcolor[RGB]{254,254,254}\makebox[2.8em][c]{0.000} &
\cellcolor[RGB]{253,252,252}\makebox[2.8em][c]{-0.001} &
\cellcolor[RGB]{250,250,253}\makebox[2.8em][c]{-0.002} &
\cellcolor[RGB]{255,255,255}\makebox[2.8em][c]{0}
\end{array}\right]
\]

\[
C =
\begin{pmatrix}
0.7294 & -0.0023 & 0.0015 \\
0.0007 & -0.9117 & -0.0007 \\
0.0055 & -0.0068 & -0.8157
\end{pmatrix}
\pm 0.0002 \quad
B^{+} =
\begin{pmatrix}
0.0054 \\
-0.0028 \\
0.0074
\end{pmatrix}
\pm 0.0001 \quad
B^{-} =
\begin{pmatrix}
-0.0034 \\
-0.0032 \\
0.0055
\end{pmatrix}
\pm 0.0001
\]

\vspace{0.5cm}

\[
\begin{array}{l}
\boldsymbol{e^{+}e^{-} \rightarrow \tau^{+}\tau^{-}} \textbf{(Z-pole)}\\
(\sqrt{s}=91.2~\text{GeV}): \\
\left|\cos\!\left(\Theta_{\tau}^{\tau^{+}\tau^{-}\mathrm{RF}}\right)\right| \leq 0.4
\end{array} \hfill
\rho_{\tau^{+}\tau^{-}} =
\left[\begin{array}{cccc}
\cellcolor[RGB]{204,138,120}\makebox[2.8em][c]{0.392} &
\cellcolor[RGB]{255,255,255}\makebox[2.8em][c]{0.001} &
\cellcolor[RGB]{255,255,255}\makebox[2.8em][c]{0.001} &
\cellcolor[RGB]{192,110,88}\makebox[2.8em][c]{0.433} \\[2pt]
\cellcolor[RGB]{255,255,255}\makebox[2.8em][c]{0.001} &
\cellcolor[RGB]{255,254,254}\makebox[2.8em][c]{-0.001} &
\cellcolor[RGB]{254,253,253}\makebox[2.8em][c]{0.003} &
\cellcolor[RGB]{254,253,253}\makebox[2.8em][c]{0.003} \\[2pt] 
\cellcolor[RGB]{255,255,255}\makebox[2.8em][c]{0.001} &
\cellcolor[RGB]{254,253,253}\makebox[2.8em][c]{0.003} &
\cellcolor[RGB]{254,254,255}\makebox[2.8em][c]{-0.002} &
\cellcolor[RGB]{254,253,253}\makebox[2.8em][c]{0.002} \\[2pt]
\cellcolor[RGB]{192,110,88}\makebox[2.8em][c]{0.433} &
\cellcolor[RGB]{254,253,253}\makebox[2.8em][c]{0.003} &
\cellcolor[RGB]{254,253,253}\makebox[2.8em][c]{0.002} &
\cellcolor[RGB]{179,100,79}\color{white}\makebox[2.8em][c]{0.610}
\end{array}\right]
+i\,
\left[\begin{array}{cccc}
\cellcolor[RGB]{255,255,255}\makebox[2.8em][c]{0} &
\cellcolor[RGB]{254,253,253}\makebox[2.8em][c]{0.001} &
\cellcolor[RGB]{254,254,254}\makebox[2.8em][c]{0.000} &
\cellcolor[RGB]{244,244,250}\makebox[2.8em][c]{-0.007} \\[2pt]
\cellcolor[RGB]{254,254,254}\makebox[2.8em][c]{-0.001} &
\cellcolor[RGB]{255,255,255}\makebox[2.8em][c]{0} &
\cellcolor[RGB]{253,252,252}\makebox[2.8em][c]{0.001} &
\cellcolor[RGB]{248,248,252}\makebox[2.8em][c]{-0.003} \\[2pt]
\cellcolor[RGB]{254,254,254}\makebox[2.8em][c]{-0.000} &
\cellcolor[RGB]{253,253,253}\makebox[2.8em][c]{-0.001} &
\cellcolor[RGB]{255,255,255}\makebox[2.8em][c]{0} &
\cellcolor[RGB]{254,253,253}\makebox[2.8em][c]{-0.001} \\[2pt]
\cellcolor[RGB]{245,244,243}\makebox[2.8em][c]{0.007} &
\cellcolor[RGB]{253,252,252}\makebox[2.8em][c]{0.003} &
\cellcolor[RGB]{254,254,254}\makebox[2.8em][c]{0.001} &
\cellcolor[RGB]{255,255,255}\makebox[2.8em][c]{0}
\end{array}\right]
\]

\[
C =
\begin{pmatrix}
-0.8709 & -0.0161 & 0.0035 \\
-0.0115 & 0.8594 & 0.0072 \\
0.0038 & 0.0029 & -1.0047
\end{pmatrix}
\pm 0.0002 \quad
B^{+} =
\begin{pmatrix}
0.00655 \\
0.00521 \\
-0.21757
\end{pmatrix}
\pm 0.00009 \quad
B^{-} =
\begin{pmatrix}
0.00585 \\
0.00065 \\
-0.21913
\end{pmatrix}
\pm 0.00009
\]

\subsection*{\large Values for $\boldsymbol{e^{+}e^{-} \rightarrow \tau^{+} \tau^{-} \gamma}$}

\[
\begin{array}{l}
\boldsymbol{e^{+}e^{-} \rightarrow \tau^{+}\tau^{-}\gamma} \textbf{(Belle II)} \\
(\sqrt{s}=10.579~\text{GeV}): \\
\quad \left|\cos\!\left(\Theta_{\tau}^{\tau^{+}\tau^{-}\mathrm{RF}}\right)\right| \leq 0.4
\end{array} \hfill
\rho_{\tau^{+}\tau^{-}} =
\left[\begin{array}{cccc}
\cellcolor[RGB]{206,141,123}\makebox[2.8em][c]{0.383} &
\cellcolor[RGB]{255,255,255}\makebox[2.8em][c]{-0.000} &
\cellcolor[RGB]{252,250,249}\makebox[2.8em][c]{0.006} &
\cellcolor[RGB]{115,127,217}\makebox[2.8em][c]{-0.232} \\[2pt]
\cellcolor[RGB]{255,255,255}\makebox[2.8em][c]{-0.000} &
\cellcolor[RGB]{227,184,167}\makebox[2.8em][c]{0.113} &
\cellcolor[RGB]{235,199,184}\makebox[2.8em][c]{0.093} &
\cellcolor[RGB]{254,254,254}\makebox[2.8em][c]{-0.002} \\[2pt]
\cellcolor[RGB]{252,250,249}\makebox[2.8em][c]{0.006} &
\cellcolor[RGB]{235,199,184}\makebox[2.8em][c]{0.093} &
\cellcolor[RGB]{225,181,164}\makebox[2.8em][c]{0.119} &
\cellcolor[RGB]{254,254,254}\makebox[2.8em][c]{-0.002} \\[2pt]
\cellcolor[RGB]{115,127,217}\makebox[2.8em][c]{-0.232} &
\cellcolor[RGB]{254,254,254}\makebox[2.8em][c]{-0.002} &
\cellcolor[RGB]{254,254,254}\makebox[2.8em][c]{-0.002} &
\cellcolor[RGB]{205,139,122}\makebox[2.8em][c]{0.386}
\end{array}\right]
+i\,
\left[\begin{array}{cccc}
\cellcolor[RGB]{255,255,255}\makebox[2.8em][c]{0} &
\cellcolor[RGB]{235,238,248}\makebox[2.8em][c]{-0.018} &
\cellcolor[RGB]{218,221,243}\makebox[2.8em][c]{-0.020} &
\cellcolor[RGB]{248,246,245}\makebox[2.8em][c]{0.002} \\[2pt]
\cellcolor[RGB]{244,229,225}\makebox[2.8em][c]{0.018} &
\cellcolor[RGB]{255,255,255}\makebox[2.8em][c]{0} &
\cellcolor[RGB]{252,245,242}\makebox[2.8em][c]{-0.005} &
\cellcolor[RGB]{244,229,225}\makebox[2.8em][c]{0.017} \\[2pt]
\cellcolor[RGB]{237,213,207}\makebox[2.8em][c]{0.020} &
\cellcolor[RGB]{245,248,253}\makebox[2.8em][c]{0.005} &
\cellcolor[RGB]{255,255,255}\makebox[2.8em][c]{0} &
\cellcolor[RGB]{237,213,207}\makebox[2.8em][c]{0.024} \\[2pt]
\cellcolor[RGB]{250,251,253}\makebox[2.8em][c]{-0.002} &
\cellcolor[RGB]{235,238,248}\makebox[2.8em][c]{-0.017} &
\cellcolor[RGB]{218,221,243}\makebox[2.8em][c]{-0.024} &
\cellcolor[RGB]{255,255,255}\makebox[2.8em][c]{0}
\end{array}\right]
\]

\[
C =
\begin{pmatrix}
0.2788 & 0.0137 & -0.0163 \\
-0.0057 & -0.6497 & -0.0745 \\
-0.0037 & -0.0822 & -0.5366
\end{pmatrix}
\pm 0.0008 \quad
B^{+} =
\begin{pmatrix}
0.0091 \\
0.0059 \\
-0.0089
\end{pmatrix}
\pm 0.0005 \quad
B^{-} =
\begin{pmatrix}
-0.0041 \\
-0.0118 \\
0.0028
\end{pmatrix}
\pm 0.0005
\]

\[
\begin{array}{l}
\boldsymbol{e^{+}e^{-} \rightarrow \tau^{+}\tau^{-}\gamma} \textbf{(Z-pole)}\\
(\sqrt{s}=91.2~\text{GeV}): \\
\left|\cos\!\left(\Theta_{\tau}^{\tau^{+}\tau^{-}\mathrm{RF}}\right)\right| \leq 0.4
\end{array} \hfill
\rho_{\tau^{+}\tau^{-}} =
\left[\begin{array}{cccc}
\cellcolor[RGB]{186,113,91}\makebox[2.8em][c]{0.326} &
\cellcolor[RGB]{243,216,210}\makebox[2.8em][c]{0.013} &
\cellcolor[RGB]{253,253,254}\makebox[2.8em][c]{-0.004} &
\cellcolor[RGB]{221,173,156}\makebox[2.8em][c]{0.130} \\[2pt]
\cellcolor[RGB]{243,216,210}\makebox[2.8em][c]{0.013} &
\cellcolor[RGB]{231,193,177}\makebox[2.8em][c]{0.079} &
\cellcolor[RGB]{255,255,255}\makebox[2.8em][c]{-0.000} &
\cellcolor[RGB]{251,250,249}\makebox[2.8em][c]{-0.005} \\[2pt]
\cellcolor[RGB]{253,253,254}\makebox[2.8em][c]{-0.004} &
\cellcolor[RGB]{255,255,255}\makebox[2.8em][c]{-0.000} &
\cellcolor[RGB]{227,184,167}\makebox[2.8em][c]{0.091} &
\cellcolor[RGB]{235,238,248}\makebox[2.8em][c]{-0.013} \\[2pt]
\cellcolor[RGB]{221,173,156}\makebox[2.8em][c]{0.130} &
\cellcolor[RGB]{251,250,249}\makebox[2.8em][c]{-0.005} &
\cellcolor[RGB]{235,238,248}\makebox[2.8em][c]{-0.013} &
\cellcolor[RGB]{192,110,88}\makebox[2.8em][c]{0.503}
\end{array}\right]
+i\,
\left[\begin{array}{cccc}
\cellcolor[RGB]{255,255,255}\makebox[2.8em][c]{0} &
\cellcolor[RGB]{242,235,231}\makebox[2.8em][c]{-0.011} &
\cellcolor[RGB]{240,223,216}\makebox[2.8em][c]{0.014} &
\cellcolor[RGB]{243,234,230}\makebox[2.8em][c]{-0.010} \\[2pt]
\cellcolor[RGB]{239,225,219}\makebox[2.8em][c]{0.011} &
\cellcolor[RGB]{255,255,255}\makebox[2.8em][c]{0} &
\cellcolor[RGB]{250,251,253}\makebox[2.8em][c]{0.000} &
\cellcolor[RGB]{240,223,216}\makebox[2.8em][c]{-0.012} \\[2pt]
\cellcolor[RGB]{235,238,248}\makebox[2.8em][c]{-0.014} &
\cellcolor[RGB]{255,255,255}\makebox[2.8em][c]{-0.000} &
\cellcolor[RGB]{255,255,255}\makebox[2.8em][c]{0} &
\cellcolor[RGB]{248,246,245}\makebox[2.8em][c]{0.002} \\[2pt]
\cellcolor[RGB]{239,225,219}\makebox[2.8em][c]{0.010} &
\cellcolor[RGB]{235,238,248}\makebox[2.8em][c]{0.012} &
\cellcolor[RGB]{250,251,253}\makebox[2.8em][c]{-0.002} &
\cellcolor[RGB]{255,255,255}\makebox[2.8em][c]{0}
\end{array}\right]
\]

\[
C =
\begin{pmatrix}
-0.2592 & -0.0209 & -0.0035 \\
-0.0194 & 0.2599 & 0.0513 \\
-0.0525 & -0.0253 & -0.6597
\end{pmatrix}
\pm 0.001 \quad
B^{+} =
\begin{pmatrix}
-0.0185 \\
-0.0041 \\
-0.1892
\end{pmatrix}
\pm 0.0006 \quad
B^{-} =
\begin{pmatrix}
0.0007 \\
0.0187 \\
-0.1649
\end{pmatrix}
\pm 0.0006
\]
\vspace{0.5cm}

\begingroup
\def\arraystretch{1.25}
\begin{table}[!h] \leftskip-1.5em
\begin{tabular}{|c|c|cc|c|c|cc|cc|cc|cc|c|}
\hline
\multirow{2}{*}{Machine} & $\sqrt{s}$ & \multicolumn{2}{c|}{$\sigma   (pb) \footnote{ Restriction to $|\cos\Theta_{\tau^{-}}^{\tau^+ \tau^- RF}|<0.4$}$} & $\int \mathcal{L}dt$ & \multirow{1}{*}{Final} & \multicolumn{2}{c|}{events   ($10^6$)} & \multicolumn{1}{c|}{$D_{n, r}$} & $\Delta D_{n, r}$ & \multicolumn{1}{c|}{$D_{n, r}$} & $\Delta D_{n, r}$ & \multicolumn{2}{c|}{$R_{D_{n, r}} = D_{n, r}^{\tau\tau\gamma} / D_{n, r}^{   \tau\tau}$} & \multirow{2}{*}{$\mathcal{S}$} \\ \cline{3-4} \cline{7-14}
 & [GeV] & \multicolumn{1}{c|}{$\tau^+\tau^-$} & $\tau^+\tau^-\gamma$ & [ab$^{-1}$] & \multirow{1}{*}{state}  & \multicolumn{1}{c|}{$\tau^+\tau^-$} & $\tau^+\tau^-\gamma$ & \multicolumn{2}{c|}{$\tau^+\tau^-$} & \multicolumn{2}{c|}{$\tau^+\tau^-\gamma$} & \multicolumn{1}{c|}{$R_{D_{n, r}}$} & $\Delta R_{D_{n, r}}$ &  \\ \hline
Belle II 2025~\cite{Belle-II:2024vuc} & \multirow{2}{*}{10.6} & \multicolumn{1}{c|}{\multirow{2}{*}{$2.96$}} & \multirow{2}{*}{$0.04$} & 0.5 & \multirow{4}{*}{$\pi^+ \pi^-$} & \multicolumn{1}{c|}{$1.48$} & $0.02$ & \multicolumn{1}{c|}{\multirow{2}{*}{$-0.819$}} & $0.001$ & \multicolumn{1}{c|}{\multirow{2}{*}{$-0.488$ \footnote{$p_T^{\gamma} > 2$ GeV}}} & $0.011$ & \multicolumn{1}{c|}{\multirow{2}{*}{$0.60$}} & $0.01$ & $34\,\sigma$ \\ \cline{1-1} \cline{5-5} \cline{7-8} \cline{10-10} \cline{12-12} \cline{14-15} 
Belle II target~\cite{Belle-II:2018jsg} &  & \multicolumn{1}{c|}{} &  & 50 &  & \multicolumn{1}{c|}{$148$} & $2.14$ & \multicolumn{1}{c|}{} & $0.0001$ & \multicolumn{1}{c|}{} & $0.001$ & \multicolumn{1}{c|}{} & $0.001$ & $290\,\sigma$ \\  \cline{1-5} \cline{7-15}

GigaZ~\cite{LinearColliderVision:2025hlt} & \multirow{2}{*}{91.2} & \multicolumn{1}{c|}{\multirow{2}{*}{$7.20$}} & \multirow{2}{*}{$0.03$} & 0.1 & & \multicolumn{1}{c|}{$0.72$} & $< 0.01$ & \multicolumn{1}{c|}{\multirow{2}{*}{$-0.912$}} & $0.002$ & \multicolumn{1}{c|}{\multirow{2}{*}{$-0.393$ \footnote{$p_T^{\gamma} > 30$ GeV}}} & $0.032$ & \multicolumn{1}{c|}{\multirow{2}{*}{$0.43$}} & $0.04$ & $16\,\sigma$ \\ \cline{1-1} \cline{5-5} \cline{7-8} \cline{10-10} \cline{12-12} \cline{14-15} 
TeraZ~\cite{FCC:2025lpp,CEPCStudyGroup:2018ghi} &  & \multicolumn{1}{c|}{} &  & 300 &  & \multicolumn{1}{c|}{$216$} & $0.86$ & \multicolumn{1}{c|}{} & $0.00003$ & \multicolumn{1}{c|}{} & $0.0006$ & \multicolumn{1}{c|}{} & $0.0006$ & $900\,\sigma$ \\ \hline
\end{tabular}
\caption{Prospected statistical significance for the loss of entanglement between $\tau^+ \tau^-$  and $\tau^+ \tau^- + \gamma$ production from $e^+e^-$ collisions, considering a single-prong final state. $D_{n, r}$ marker predictions are provided for the inclusive $t\bar{t}$ sample and with an optimized $p_T^\gamma$ cut for $\tau^+ \tau^- + \gamma$. The $D_n$ marker is still used in Belle II prospects. Note that, for Z-pole prospects, the decoherence observables is the $D_r$ marker.}
\end{table}
\endgroup

\newpage

\end{document}